\documentclass[twocolumn,aps,prd,superscriptaddress,notitlepage,10pt,final,longbibliography,nofootinbib]{revtex4-1}

\usepackage[pdftex]{graphicx}
\usepackage[pdftex]{xcolor}
\usepackage[pdftex]{hyperref}

\usepackage{float}

\usepackage{amsmath,amssymb,mathrsfs}

\usepackage{subfigure}
\usepackage{epsf}
\usepackage{bbm}
\usepackage{color}
\usepackage{cleveref}
\usepackage{comment}
\usepackage{natbib}
\usepackage{appendix}
\usepackage{here}

\graphicspath{{./fig_arxiv/}}

\begin{document}
\title{Multishell Dirac fermions in the Einstein-Dirac system}
	
\author{Shunsuke Iwasawa~}
\email{iwasawa.phys.jazz@gmail.com}
\affiliation{Department of Physics and Astronomy, Tokyo University of Science, Noda, Chiba 278-8510, Japan}

\author{Nobuyuki Sawado~}
\email{sawadoph@rs.tus.ac.jp}
\affiliation{Department of Physics and Astronomy, Tokyo University of Science, Noda, Chiba 278-8510, Japan}

\begin{abstract}
We present multifermions in the spherically symmetric Einstein-Dirac system. 
Dirac fermions are self-localized within a spherically symmetric Einstein gravity, i.e., 
the Schwarzschild-like space-time metric. 
Most of previous studies of the Einstein-Dirac system are restricted to two 
neutral fermions or to many fermions with the same high-angular momentum filling a single shell. 
Our model considers full-filling of fermions in multiple shells, similarly to the 
conventional nuclear shell model. 
We solve the model for fermion numbers $N_\textrm{f}=2,6,12$ and $20$, which 
can realize a spherically symmetric system. 
Even single-shell multifermions exhibit a multipeak structure and 
fragmentation in the high redshift region. 
The behavior observed in our multishell model can be
explained by interactions between the shells and resulting delocalization. 
We also investigate the pressure of the solutions which defines the existence (or absence) of 
intershell interactions. 
The radial pressure is attractive, supporting compactness of the solutions. 
Finally, we show the correlations between the nontrivial changes in the Shannon entropy 
(a logarithmic measure of information content) 
and the structural deformations of the solutions. 
\end{abstract}

\maketitle

\section{Introduction}

The Einstein-Dirac system is a system of fermions minimally coupled with Einstein gravity~\cite{Deser:1974cy}
in which the fermions localize on finite region like solitons.
Since Finster \textit{et al}.~\cite{Finster:1998ws} explicitly solved the system in spherically symmetric space-time, 
many types of solutions have been discussed. 
The Pauli exclusion principle prevents the collapse of Dirac spinor configurations under their own gravitational fields. 

The literature on the Einstein-Dirac system has discussed noninteracting free Dirac fermions and
nonlinear self-interactions as the four-point forms
~\cite{Bronnikov:2019nqa,Dzhunushaliev:2018jhj,daRocha:2020rda}. 
Numerous studies have investigated the Dirac fermions in gravitating space-time,
coupling of Dirac fermions with electromagnetism~\cite{Finster:1998jqw}, and  
Yang-Mills fields~\cite{Finster:2000vy} (see also recent follow-up studies  
in~\cite{Dzhunushaliev:2019kiy,Dzhunushaliev:2019uft,Blazquez-Salcedo:2020czn,Kain:2023ann,
Kain:2023pvp,Dzhunushaliev:2023ylf}.) 
Localized fermions have also been coupled with gravitating-skyrmions~\cite{Dzhunushaliev:2024kti}. 
The typical spectral flow of the fermions in such coupled system has been extensively studied. 
The properties of Dirac fermions on baby-skyrmions 
in anti-de Sitter space-time have been thoroughly discussed in the context of six-dimensional
brane-worlds~\cite{Kodama:2008xm,Delsate:2011aa}.
Some analytical solutions of fermions in five-dimensional warped space-time 
with special potentials have shown in~\cite{Cui:2022djf,Zhou:2025dyr}.

In this paper, we study the localization of many fermion states in a spherically symmetric, 
Schwarzschild-like self-gravitating system. 
In the spherically symmetric case, 
we can discuss at least two fermions. 
In the solutions of~\cite{Finster:1998ws}, spherical symmetry is realized when 
the fermions have opposite spins. 
This setup has been extended by many authors~\cite{Herdeiro:2017fhv,Bronnikov:2019nqa}. 
Recently, Leith \textit{et al}.~\cite{Leith:2020jqw,Leith:2021urf} extended the formulation
of~\cite{Finster:1998ws} to many fermion systems.   
The fermions, numbering $N_\textrm{f}=2j+1$ are arranged in a filled shell with a total
angular momentum taken from $j\in \{\frac{1}{2},\frac{3}{2},\frac{5}{2},\cdots\}$,  
retaining spherical symmetry of the system. 
This model prominently features a multipeak structure and
fragmentation especially in the high-redshift region~\cite{Leith:2020jqw,BakuczCanario:2020qmq}. 
A concrete interpretation of this phenomenon is currently lacking. 
As the shell's interior contains a large number of unoccupied states, the solutions appear 
to be excited states of the system. 
The authors of Ref.\cite{Andreasson:2025uir} discussed the similarity between the solutions of
the Einstein-Dirac system and the classical Einstein-Vlasov model. 
Multiple fermions have been handled by various approaches, 
such as the multiple spinor formulation~\cite{Blazquez-Salcedo:2019uqq,Liang:2023ywv}, 
and two possible shells with $j=\frac{1}{2}$ and $\frac{3}{2}$ within context of
semiclassical gravity~\cite{Kain:2023jgu}. 
Although Finster~\cite{finster:2002} studied their complete multifermion shells for discussing the absence of black hole 
solutions in the Einstein-Dirac-Yang/Mills system, they did not provide explicit solutions.
Here we aim to expand the number of shells, providing three- or four-shell solutions. 
We study the energy or eigenstates over a wide range of redshift 
parameter~\footnote{Leith \textit{et al}.~\cite{Leith:2020jqw} defined the central redshift as 
$z\equiv T_0-1$. We borrow their definitions with some simplification.} $T_0$. 
We also investigate the structural changes of the shells in detail,  
including shell fragmentation~\cite{Leith:2020jqw}, pressure~\cite{Andreasson:2025uir}, 
and Shannon's informational entropy \cite{Gleiser:2011di}.  
The investigated system possesses several 
excited nodal solutions that support negative-parity states~\cite{Sporea:2019iwk,Leith:2021urf,Kain:2023jgu}.  
Our multishell model is expected to yield a vast number of solutions.  
These solutions can be straightforwardly investigated but
but such comprehensive results are not required for understanding the basic features of the system. 
Therefore, this paper analyzes only the ground state (non-nodal) solutions. 

The remainder of this paper is organized as follows. 
Section \ref{sec:2} briefly introduce the Einstein-Dirac system and derives the Schwarzschild-like metric, 
the gamma matrices and the coupled Dirac and Einstein equations. 
The numerical method is overviewed in Sec.~\ref{sec:3}. 
Section \ref{sec:4} provides successful solutions and energies of the system.
Phase diagrams of the mass-energy, mass-eigenvalues, and mass-radius are presented here. 
Section \ref{sec:5} analyzes shell fragmentation in the system and 
discusses Shannon's information entropy.  
Concluding remarks are presented in Sec.~\ref{sec:6}.

\section{\label{sec:2}~Einstein-Dirac system}

This section provides the basic formulation of the Einstein-Dirac system, which 
has been extensively studied in recent years. 
Throughout this paper, we adopt a mostly positive $(-,+,+,+)$ metric signature.  
The action of the Einstein-Dirac system is given by
\begin{align}
&S_{D}=\int\left(\frac{1}{8\pi G}\mathcal{R}
+\mathcal{L}_{\mathrm{D}}\right)\sqrt{-\mathrm{det}(g_{\mu\nu})}d^4x
\label{EDaction}
\end{align}
where $\mathcal{R}$ is the Ricci scalar, $G$ is the gravitational constant, $\mathcal{L}_\mathcal{D}$ 
is the Dirac Lagrangian and $g_{\mu\nu}$ is the metric of the space-time. 
The Lagrangian of the Dirac spinor $\Psi$ is minimally coupled to gravity 
as follows:  
\begin{align}
&\mathcal{L}_{\mathrm{D}}=\bar{\Psi}\left(\mathfrak{D}-m\right)\Psi,
\end{align}
where $m$ is the fermion mass and 
$\mathfrak{D}:=i\gamma^\mu(\partial_\mu-\Gamma_\mu)$ is the Dirac operator 
in curved space-time. Here, 
$\Gamma_\mu$ defines the spin connection and $\gamma^\mu$ are the gamma matrices
in curved space-time, 
which satisfy $\{\gamma^\mu,\gamma^\nu\}=-2g^{\mu\nu}$ in Clifford algebra.  
The variation of action \eqref{EDaction} with respect to the metric and Dirac field yields the 
Einstein and the Dirac equations
\begin{align}
	&G_{\mu\nu}=R_{\mu\nu}-\frac12g_{\mu\nu}R=8\pi GT_{\mu\nu},
	\label{formEinsteineq}\\
	&\left(\mathfrak{D}-m\right)\Psi=0.
	\label{formDiraceq}
\end{align}
The energy-momentum tensor $T_{\mu\nu}$ in Eq.\eqref{formEinsteineq} is derived 
in terms of the Dirac Lagrangian as
\begin{align}
	&T_{\mu\nu}=\frac{\delta \mathcal{L}_{\mathrm{D}}}{\delta g^{\mu\nu}}-\frac12g_{\mu\nu} \mathcal{L}_{\mathrm{D}}\,.	
	\label{EMtensor}
\end{align}
When a shell is fully filled with fermions, the object will be spherically symmetric. 
Therefore, we solve the coupled equations \eqref{formEinsteineq}-\eqref{formDiraceq} 
in spherical coordinates $(t,r,\theta,\phi)$, using
the spherically symmetric metric
\begin{align}
g_{\mu\nu}&=\mathrm{diag}\left(-\frac1{T(r)^2},~\frac1{A(r)},~r^2,~r^2\sin^2\theta\right)\,.
\end{align}
The fields $T(r),A(r)$ asymptotically connect to the well-known 
Schwarzschild metric as follows:
\begin{align}
T(r)^{-2},A(r)\to 1-\frac{2GM}{r}
\end{align}
where $M$ is the Arnowitt-Deser-Misner mass. 

Finster \textit{et al.}~\cite{Finster:1998ws,Finster:1998jqw} numerically solved  
the system~\eqref{formEinsteineq}--\eqref{formDiraceq} for two gravitationally localized neutral fermions, 
assuming opposite spins to ensure spherical symmetry. 
Leith \textit{et al.}\cite{Leith:2020jqw,Leith:2021urf} recently extended this formulation to  
many fermion systems, arranging $N_\textrm{f}$th fermions 
in a filled shell of total angular momentum $j=\frac{N_\textrm{f}-1}{2}$. 
This arrangement retains the simplifications of spherical symmetry. 
A system with an impressive spatial structure, 
including frequent density oscillations, can also be constructed.  

In the present paper, we denote the states of the total angular momentum $j_n:=\frac{2n-1}{2}$ by
$\{n\}$-shell and write
$\{1,2,\cdots,n\}$-shell for multishell solutions; correspondingly, we write 
the angular momenta of paired shells as $\{\frac{1}{2},\frac{3}{2},\cdots,\frac{2n-1}{2}\}$.  
The spinor of the $\{n\}$-shell in spherical symmetry labeled by the $n$th angular momentum and the 
$z$-component $k_n$ can be defined as the Hartree-Fock product
of the $(2j_n+1)$th state
\begin{align}
\Psi_{j_n}:=\Psi_{j_n,k_n=-j_n}\land\Psi_{j_n,k_n=-j_n+1}\land\cdots\land\Psi_{j_n,k_n=j_n}.
\label{Idvspinor}
\end{align}
The spinor wave function of an individual fermion is defined as 
\begin{align}
		&\Psi_{j_n,k_n}=e^{- i\omega_n t}\frac{\sqrt{T(r)}}{r}
		\begin{pmatrix}	
		\mathcal{Y}^{\ell}_{j_nk_n}(\theta,\phi)\alpha_n(r)\\
		i\mathcal{Y}^{\ell'}_{j_nk_n}(\theta,\phi)\beta_n(r)\\
		\end{pmatrix} 
\end{align}
where $\omega_n$ is the eigenvalue and $\ell=k_n-\frac{1}{2}, \ell'=k_n+\frac{1}{2}$. 
The angular component of the wave function $\mathcal{Y}^{\ell}_{j_nk_n}$ with 
total angular momentum and $z$-component $k_n$,
orbital angular momentum $\ell$ and spin $S=\frac{1}{2}$, is defined as  
\begin{align}
		&\mathcal{Y}^{\ell}_{j_nk_n}=\sum_{m_s}
		\left(\ell,\frac{1}{2};k_n-m_s, m_s \bigg| \ell,\frac{1}{2};j_s,k_n\right)
		\nonumber \\
		&\hspace{2cm}\times Y_{\ell k_n-m_s}(\theta,\phi)\chi_{m_s}\,.
\end{align}
where $Y_{\ell k_n-m_s}(\theta,\phi)$ define the well-known spherical harmonics. 
Employing the explicit form of the spinor $\chi_{m_s}$
\begin{align}
		&\chi_{\frac12}=
		\begin{pmatrix}
		1\\
		0\\
		\end{pmatrix},~~~~
		\chi_{-\frac12}=
		\begin{pmatrix}
		0\\
		1\\
		\end{pmatrix}\,,
\end{align}
we can explicitly write the angular part of the spinor $\mathcal{Y}$ as
\begin{align}
&\mathcal{Y}^{\ell}_{j_nk_n}=
\begin{pmatrix}
\sqrt{\frac{j_n+k_n}{2j_n}}Y_{\ell k_n-\frac12}\\
\sqrt{\frac{j_n-k_n}{2j_n}}Y_{\ell k_n+\frac12}\\
\end{pmatrix}\,,~~~~\ell=j_n-\frac12\,,
\\
&\mathcal{Y}^{\ell'}_{j_nk_n}=
\begin{pmatrix}
\sqrt{\frac{j_n+1-k_n}{2j_n+2}}Y_{\ell' k_n-\frac12}\\
-\sqrt{\frac{j_n+1+k_n}{2j_n+2}}Y_{\ell' k_n+\frac12}\\
\end{pmatrix}\,,~~~~\ell'=j_n+\frac12\,.
\end{align}
As our study is limited to static and spherically symmetric solutions, 
we can separate the fermion wave functions in a shell, where each fermion possesses the 
same energy $\omega_n$ and the radial functions differ only in their angular dependences.

The $\gamma$-matrices in curved space-time $\gamma^\mu$ are related to normal (flat) $\gamma$-matrices
$\overline{\gamma}^\mu$ as follows:
\begin{align}
\gamma^\mu=e_a^\mu\overline{\gamma}^a
\end{align}
where the vierbein $e_a^\mu$ is defined by
\begin{align}
g_{\mu\nu}=e_{a\mu}e^a_\nu,~~~\eta_{ab}=e_{a\mu}e_b^\mu
\label{vierbein}
\end{align}
where $\eta_{ab}=\textrm{diag}(-1,1,1,1)$. 
The flat $\overline{\gamma}^\mu$ satisfy the Clifford algebra
\begin{align}
\{\overline{\gamma}^a,\overline{\gamma}^b\}=-2\eta^{ab}\,,
\end{align}
respectively. 
We employ the following Dirac representation of the $\gamma$-matrices
\begin{align}
\overline{\gamma^0}&=i
\begin{pmatrix}
1&0\\
0&-1\\
\end{pmatrix},~~
\overline{\gamma^i}&=i
\begin{pmatrix}
0 &\sigma^i \\
-\sigma^i & 0 \\
\end{pmatrix} ,~~~~i=1,2,3,
\label{gamma}
\end{align}
where $\sigma^i$ denote the standard Pauli matrices. 
Using the definition in~\eqref{vierbein}, 
the vierbein is explicitly written as
\begin{align}
&e_a^\mu=
\begin{pmatrix}
T&0&0&0\\
0&\sqrt{A}\sin\theta\cos\phi&\sqrt{A}\sin\theta\sin\phi&\sqrt{A}\cos\theta\\
0&\frac{1}{r}\cos\theta\cos\phi&\frac{1}{r}\cos\theta\sin\phi&-\frac{1}{r}\sin\theta\\
0&-\frac{1}{r\sin\theta}\sin\phi&\frac{1}{r\sin\theta}\cos\phi&0\\
\end{pmatrix} 
\nonumber \\
&\hspace{3cm}a=(0,1,2,3),\mu=(t,r,\theta,\phi)\,.
\end{align}
We can now introduce the $\gamma$-matrices in polar coordinate as:
\begin{align}
\gamma^t&=T\overline{\gamma^t}\,,\\
\gamma^r&=\sqrt{A}\left(\overline{\gamma^1}\sin\theta\cos\phi+\overline{\gamma^2}\sin\theta\sin\phi+\overline{\gamma^3}\cos\theta\right)\,.
\\
\gamma^\theta&=\frac1r\left(\overline{\gamma^1}\cos\theta\cos\phi+\overline{\gamma^2}\cos\theta\sin\phi-\overline{\gamma^3}\sin\theta\right)\,,\\
\gamma^\phi&=\frac1{r\sin\theta}\left(-\overline{\gamma^1}\sin\phi+\overline{\gamma^2}\cos\phi\right)\,.
\end{align}
The spin connections are given by
\begin{align}
\Gamma_t&=\frac{\partial_r T}{T^3}\gamma^t\gamma^r,\\
\Gamma_r&=0,\\
\Gamma_\theta&=\frac{r}{2}\left(\frac{1}{\sqrt{A}}-1\right)\gamma^\theta\gamma^r,\\
\Gamma_\phi&=\frac{r\sin^2\theta}{2}\left(\frac{1}{\sqrt{A}}-1\right)\gamma^\phi\gamma^r.
\label{spin connention}
\end{align}
Finally, the Dirac operator is written as
\begin{align}
\mathfrak{D}=i\gamma^t \frac{\partial}{\partial t}+&i\gamma^r\left(\frac{\partial}{\partial r}+\frac{1}{r}\Bigl(1-\frac1{\sqrt{A}}\Bigr)-\frac {T'}{2T}\right)\nonumber\\&+i\gamma^\theta \frac{\partial}{\partial \theta}+i\gamma^\phi \frac{\partial}{\partial \phi}.
\end{align}

Leith \textit{et al}.~\cite{Leith:2020jqw} assumed that every fermion occupies the same shell, 
while other inside shells are empty.   
This paper examines multishell models with complete occupation of fermions in each shell, 
which maintain spherical symmetry. 
Setting the outer shell as the $\{N\}$-shell, the total number of fermions is 
\begin{align}
N_{\textrm{f}}=N(N+1),~~N=1,2,\cdots\,.
\end{align}

\begin{figure*}[htbp]
	\centering 
	\includegraphics[width=1.0\linewidth]{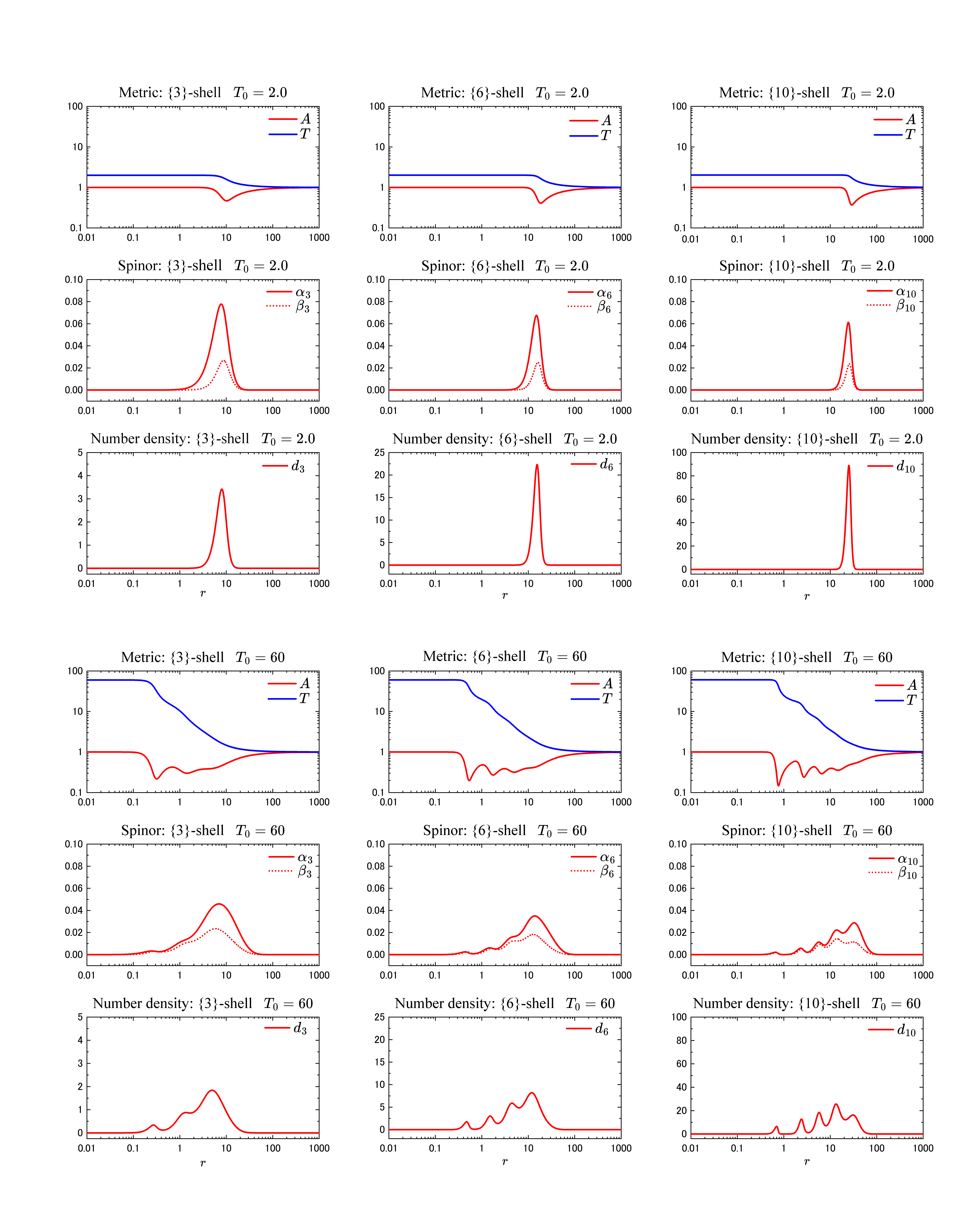}
	\caption{\label{1shell}Radial structures of the metric and fermion fields in the single-shell 
	approximation. The upper nine panels show the results at redshift $T_0=2.0$ for $N_\textrm{f}=6$ (left column), 
	$N_\textrm{f}=12$ (middle column) and $N_\textrm{f}=20$ (right column); 
	The lower nine panels show the results at $T_0=60.0$ for $N_\textrm{f}=6$ (left column), 
	$N_\textrm{f}=12$ (middle column) and $N_\textrm{f}=20$ (right column). 
	}
\end{figure*}

\begin{figure*}[htbp]
	\centering 
	\includegraphics[width=0.3\linewidth]{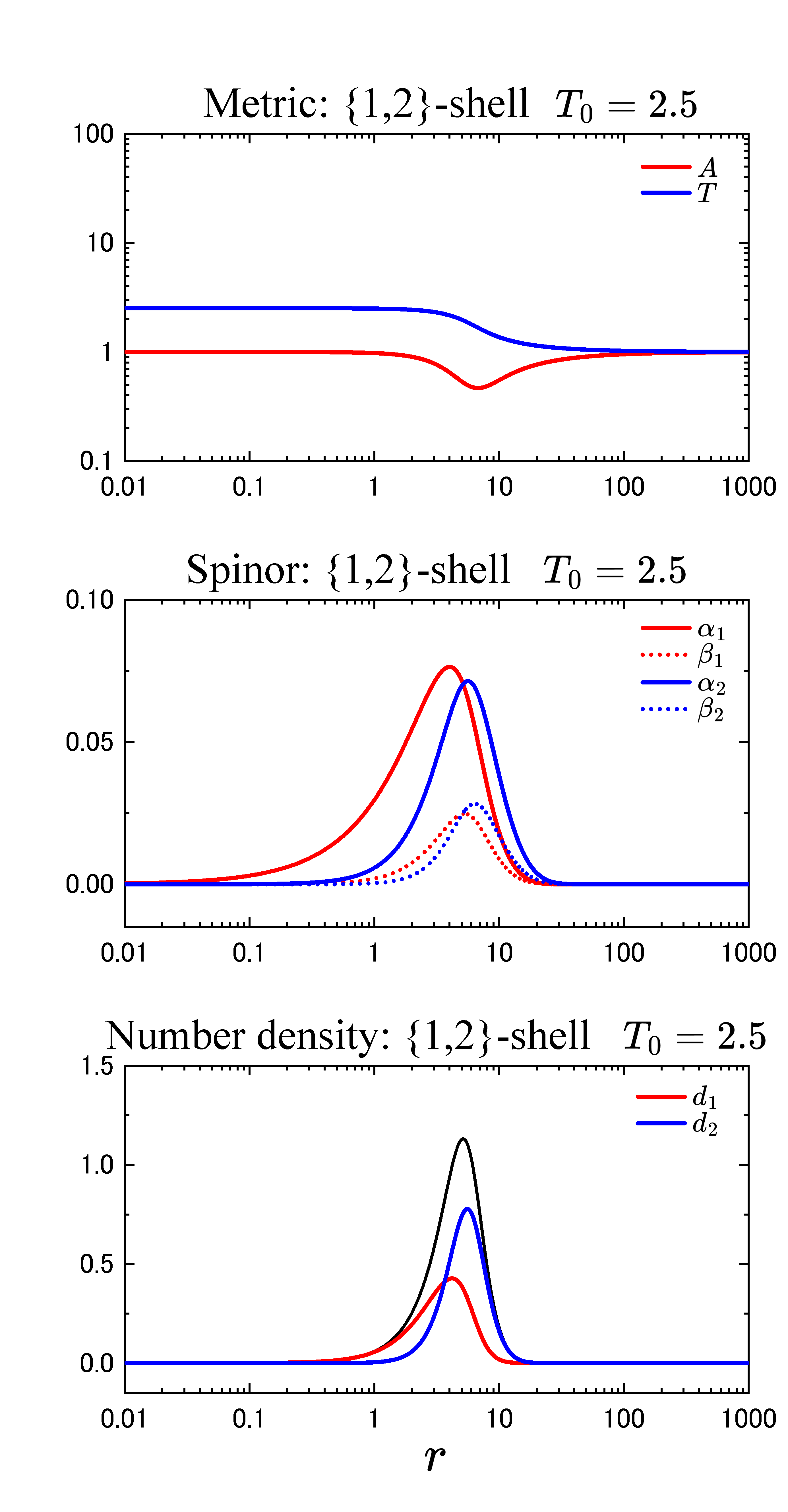}\hspace{-0.1cm}
	\includegraphics[width=0.3\linewidth]{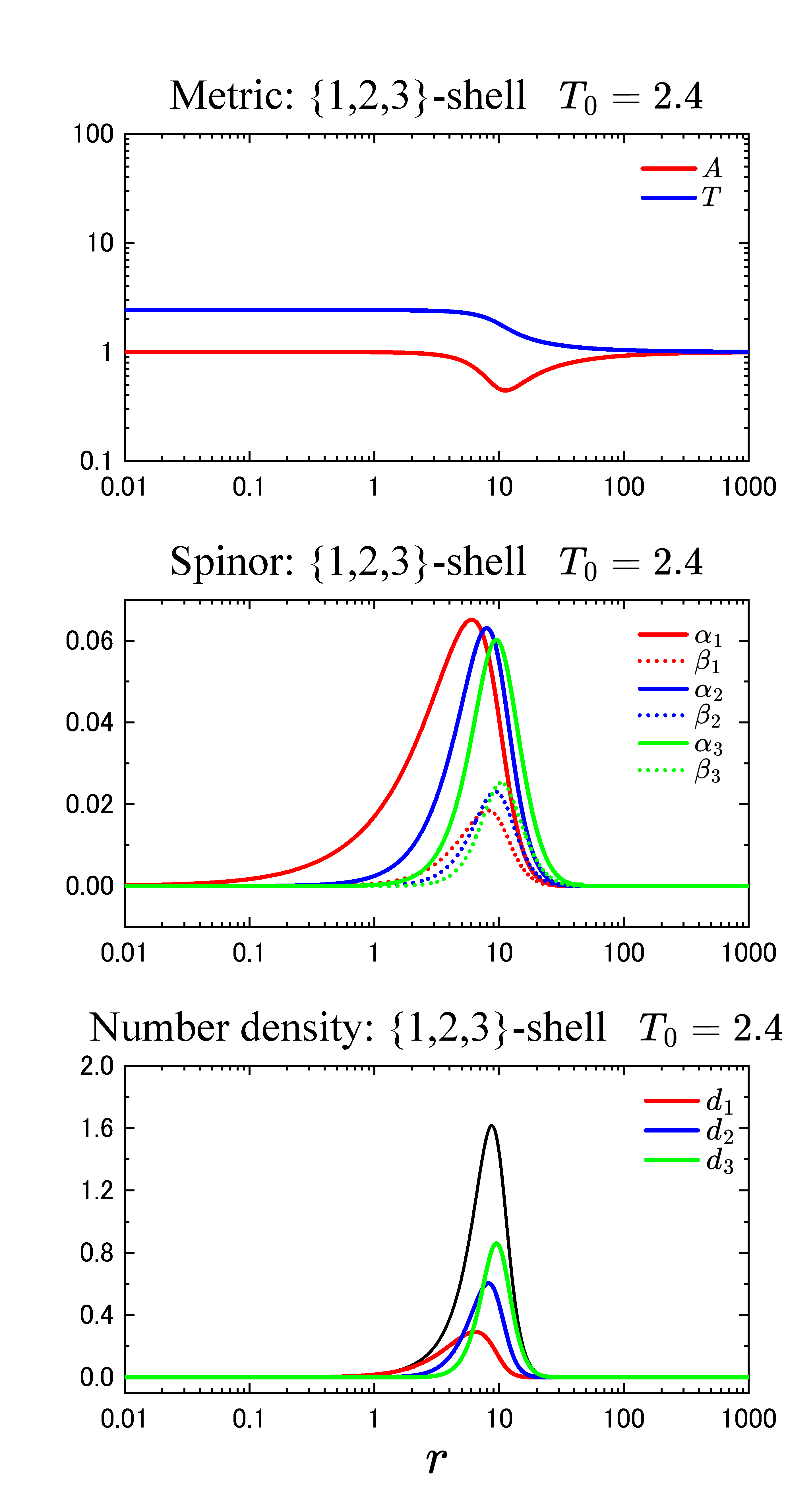}\hspace{-0.1cm}
	\includegraphics[width=0.3\linewidth]{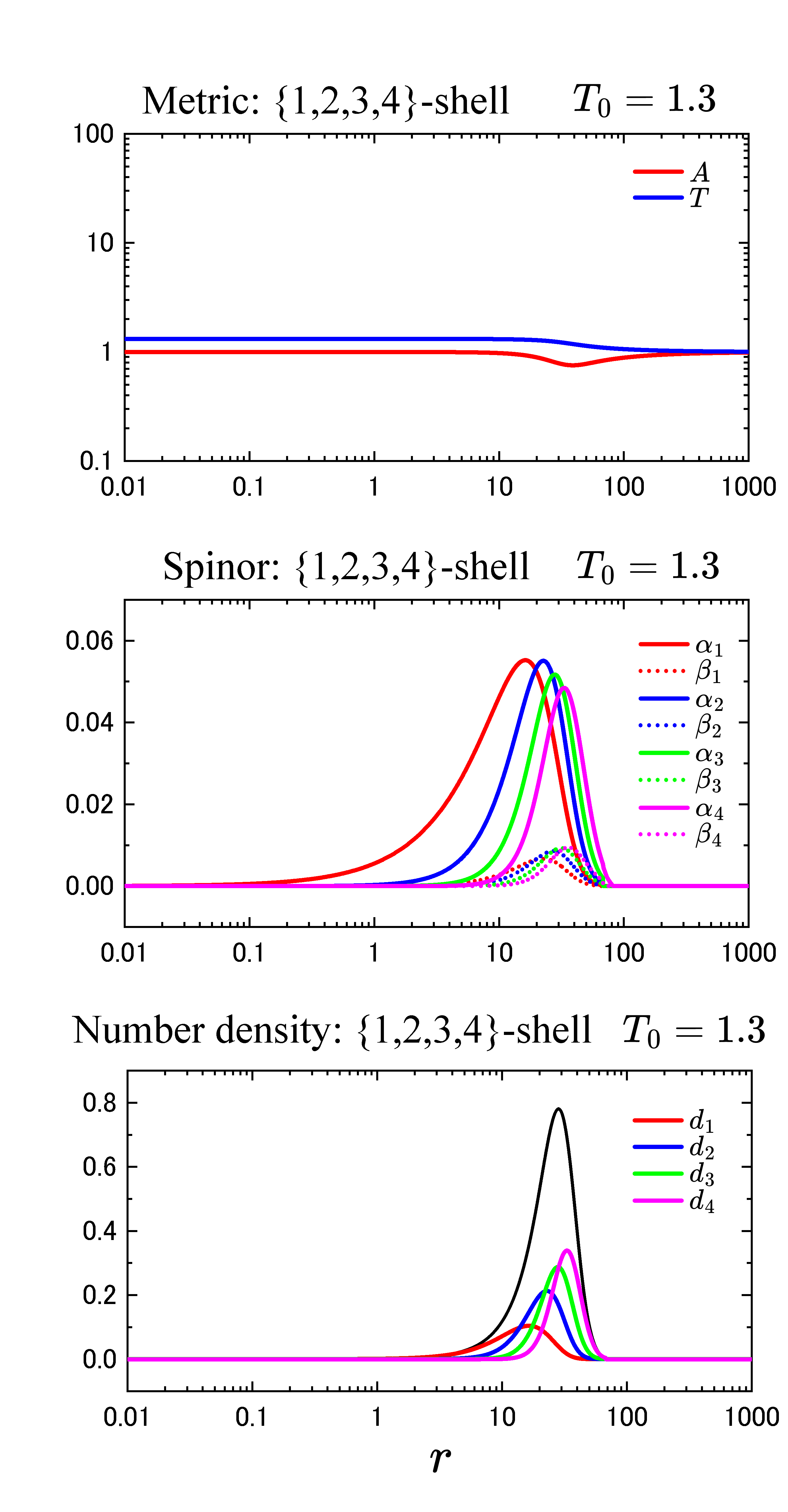}
	\includegraphics[width=0.3\linewidth]{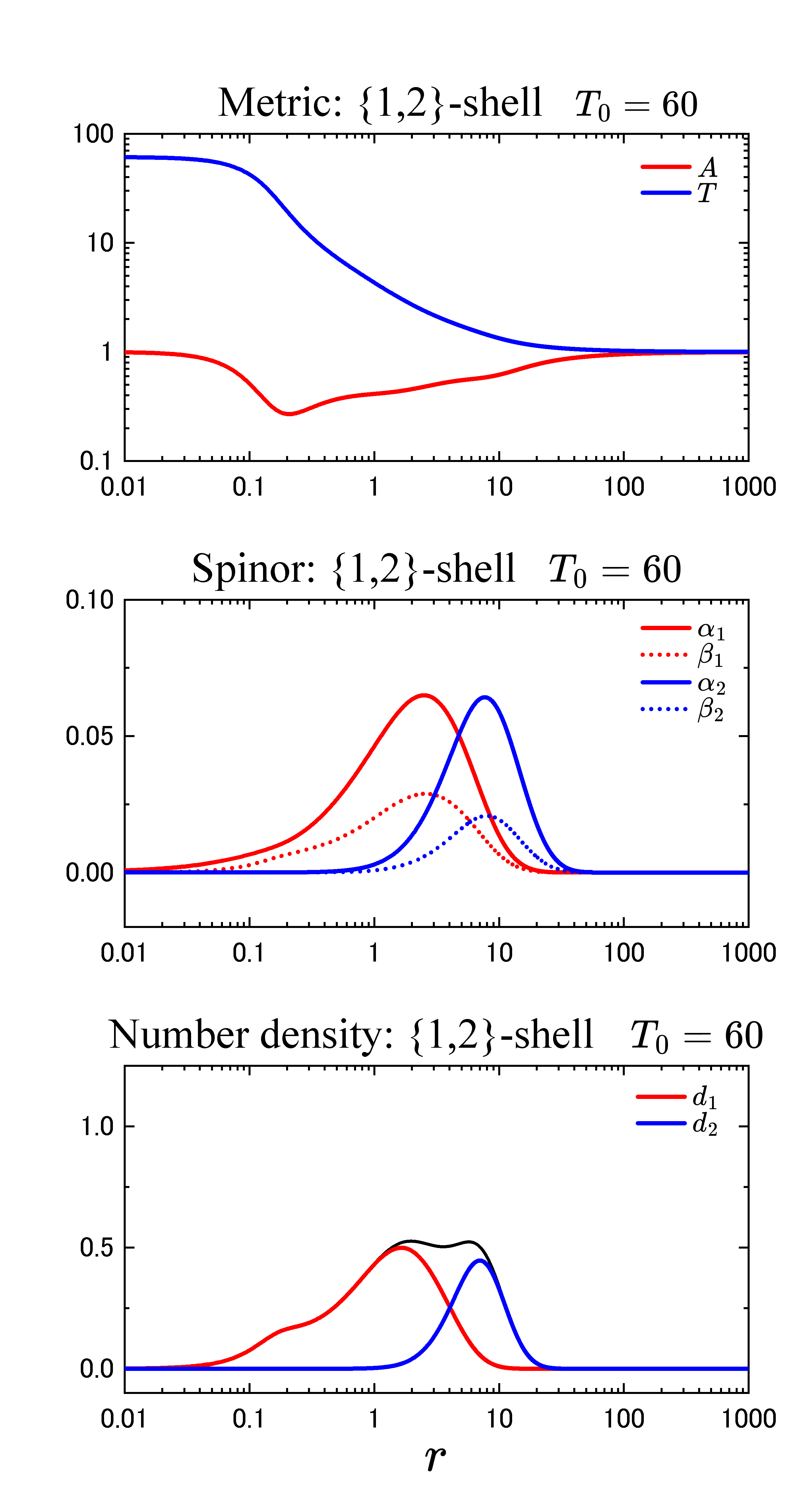}\hspace{-0.1cm}
	\includegraphics[width=0.3\linewidth]{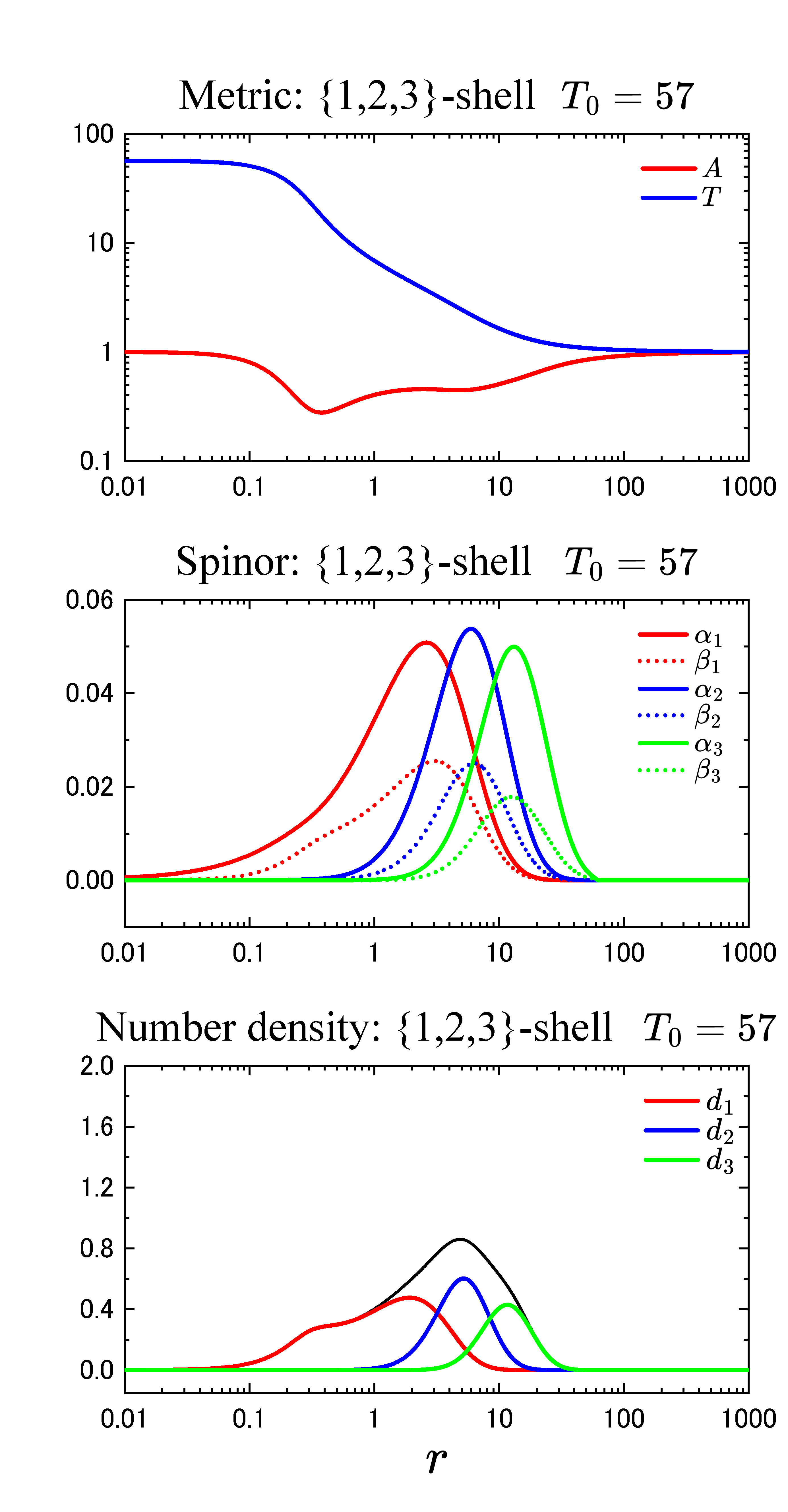}\hspace{-0.1cm}
	\includegraphics[width=0.3\linewidth]{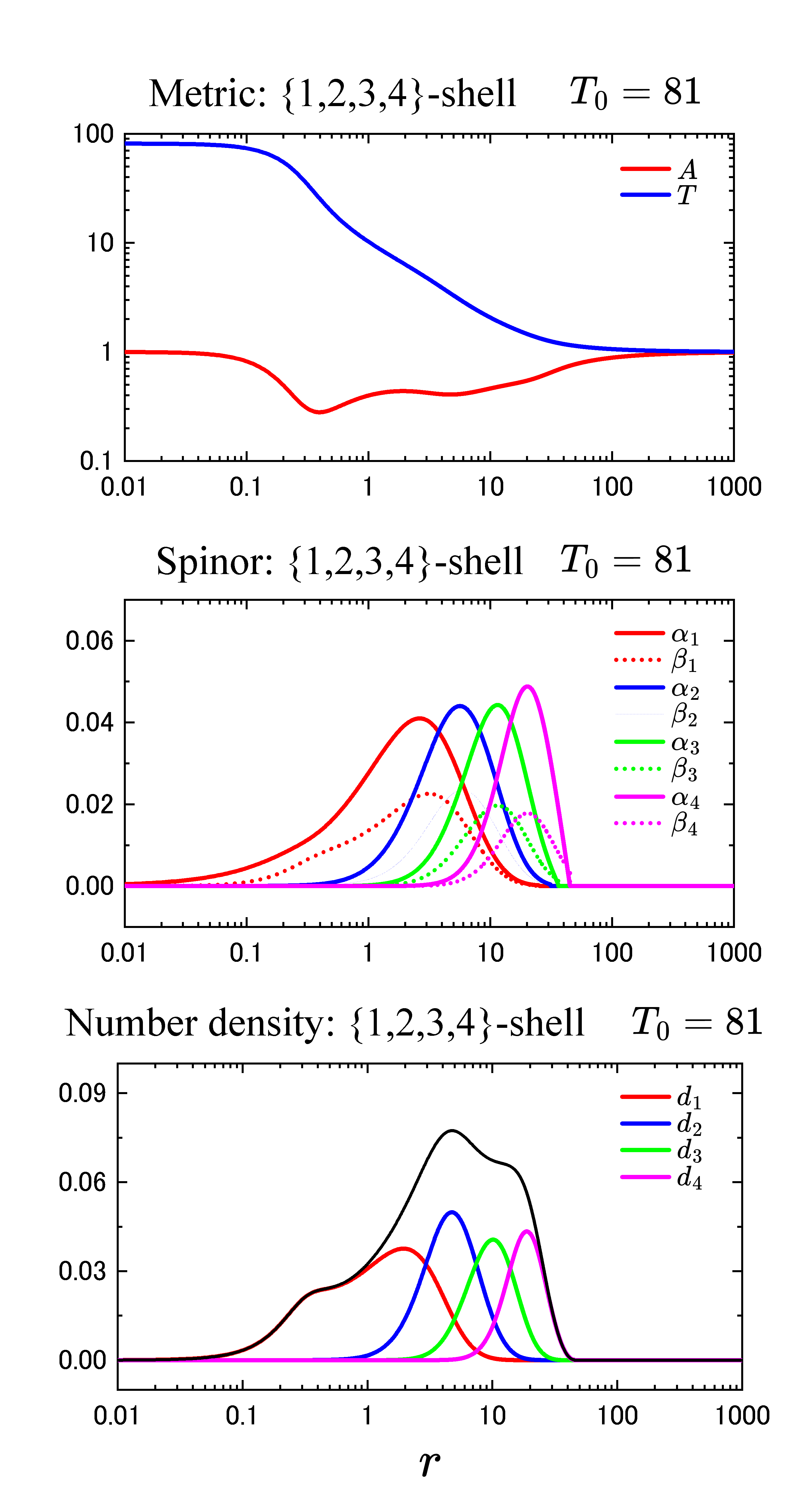}
	\caption{\label{234shell}Radial structures of the metric,
	the spinor fields, and number densities in the multishell model. 
	The black curves in the number-density plots are the
	summed densities $d=\sum_{i=1}^N d_i$.
	The upper nine panels were obtained at low redshift: $T_0=2.5$, $N_\textrm{f}=6$ (left column), 
	$T_0=2.4$, $N_\textrm{f}=12$ (middle column), and $T_0=1.3$, $N_\textrm{f}=20$ (right column) solutions. 
	The lower nine panels were obtained at high redshift: $T_0=60.0$, $N_\textrm{f}=6$ (left column), 
	$T_0=57.0$, $N_\textrm{f}=12$ (middle column), and $T_0=81.0$, $N_\textrm{f}=20$ (right column) solutions. 	
	}
\end{figure*}

\begin{table} 
\caption{\label{tab:example}The radial distribution of the number density of the single-shell and 
the multi-shell solutions corresponding to Figs.\ref{1shell}-\ref{234shell}.} 
\begin{ruledtabular} 
\begin{tabular}{p{0.2cm}p{1.2cm} l@{\hspace{-0.6em}} l @{\hspace{0.3em}}l@{\hspace{-0.6em}} l} 
$N_\textrm{f} $ &&  single-shell&$(T_0)$ & multishell&$(T_0)$ 
\\
\hline 
6 & low $T_0$ & $r$:(3.0,11.0)&(2.0) & $r$:(0.2,12.0)&(2.5) \\
& high $T_0$ & $r$:(0.2,30.0)&(60.0) & $r$:(0.02,25.0)&(60.0) \\
12 & low $T_0$ & $r$:(7.2,23.0)&(2.0) & $r$:(0.3,27.0)&(2.4) \\
& high $T_0$ & $r$:(0.3,50.0)&(60.0) & $r$:(0.02,50.0)&(57.0) \\
20 & low $T_0$ & $r$:(12.5,38.0)&(2.0) & $r$:(0.7,72.0)&(1.3) \\
& high $T_0$ & $r$:(0.5,70.0)&(60.0) & $r$:(0.02,48.0)&(81.0) \\
\end{tabular} 
\end{ruledtabular} 
\end{table}

Summing the terms in each shell, the energy-momentum tensor~\eqref{EMtensor} becomes 
\begin{align}
&T_{tt}=\sum_{n=1}^N\frac{\omega_n n}{2\pi r^2}(\alpha_n^2+\beta_n^2)\,,
\label{EMt}
\\
&T_{rr}=\frac{1}{2\pi Ar^2}\sum_{n=1}^N\biggl(\omega_n n T^2(\alpha_n^2+\beta_n^2)
\nonumber 
\\
&\hspace{2cm} -mn T(\alpha_n^2-\beta_n^2)-\frac{n^2}{2\pi r}\alpha_n\beta_n\biggr)\,
\label{EMr},
\end{align}
where we have used the relation
\begin{align}
\sum_{k_n=-j_n}^{j_n}\mathcal{Y}^{k_n\pm\frac{1}{2}*}_{j_nk_n}\mathcal{Y}^{k_n\pm\frac{1}{2}}_{j_nk_n}=\frac{n}{2\pi}
\label{Ycomplete}
\end{align}
to guarantee spherically symmetric solutions.
Using the ansatz for the metric and fermion wave functions, we can obtain the explicit 
form of the $2N$th coupled Dirac equations
\begin{align}
&\sqrt{A}\alpha_n'=\frac{n}{4\pi r}\alpha_n-(\omega_n T+m)\beta_n\label{Diraca},
\\
&\sqrt{A}\beta_n'=(\omega_n T-m)\alpha_n-\frac{n}{4\pi r}\beta_n,~n=1,2,\cdots,N\label{Diracb},		
\end{align}
and the $(tt)$ and $(rr)$ components of the Einstein equations
\begin{align}
&rA'=1-A-4 G\sum_{n=1}^N n\omega_n T^2(\alpha_n^2+\beta_n^2)\label{EinsteinA},
\\
&2rA\frac{T'}{T}=A-1-4 G\sum_{n=1}^N\biggl[n\omega_n T^2(\alpha_n^2+\beta_n^2)
\nonumber \\
&\hspace{1cm}
-\frac{n^2}{2\pi r}T\alpha_n\beta_n- n mT(\alpha_n^2-\beta_n^2)\biggr].\label{EinsteinT}
\end{align}
Note that the $(\theta\theta),(\phi\phi)$-components of the Einstein equation automatically 
satisfy the above equations.

\begin{figure*}[t]
	\centering 
	\includegraphics[width=0.3\linewidth]{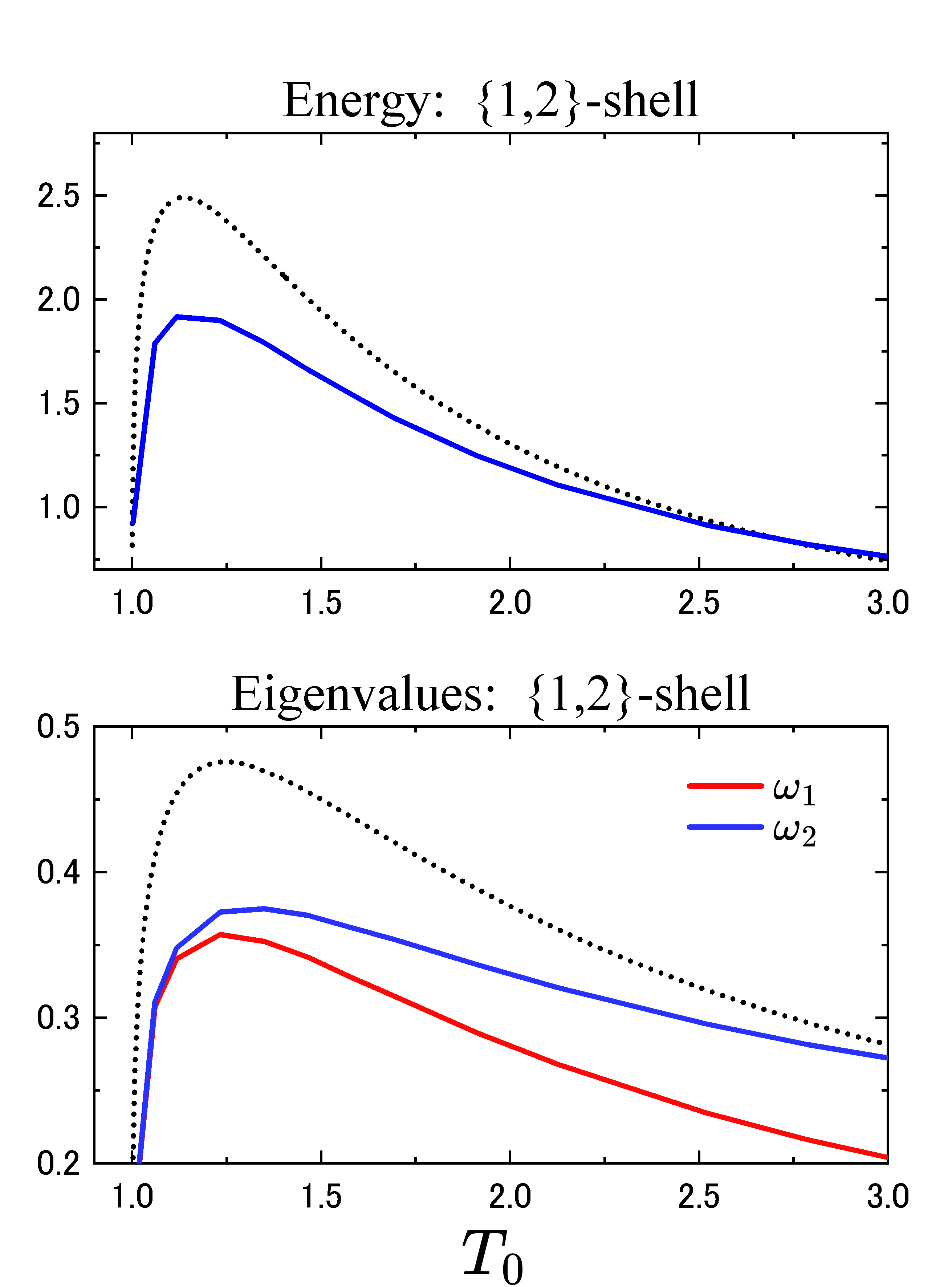}\hspace{-0.1cm}
	\includegraphics[width=0.3\linewidth]{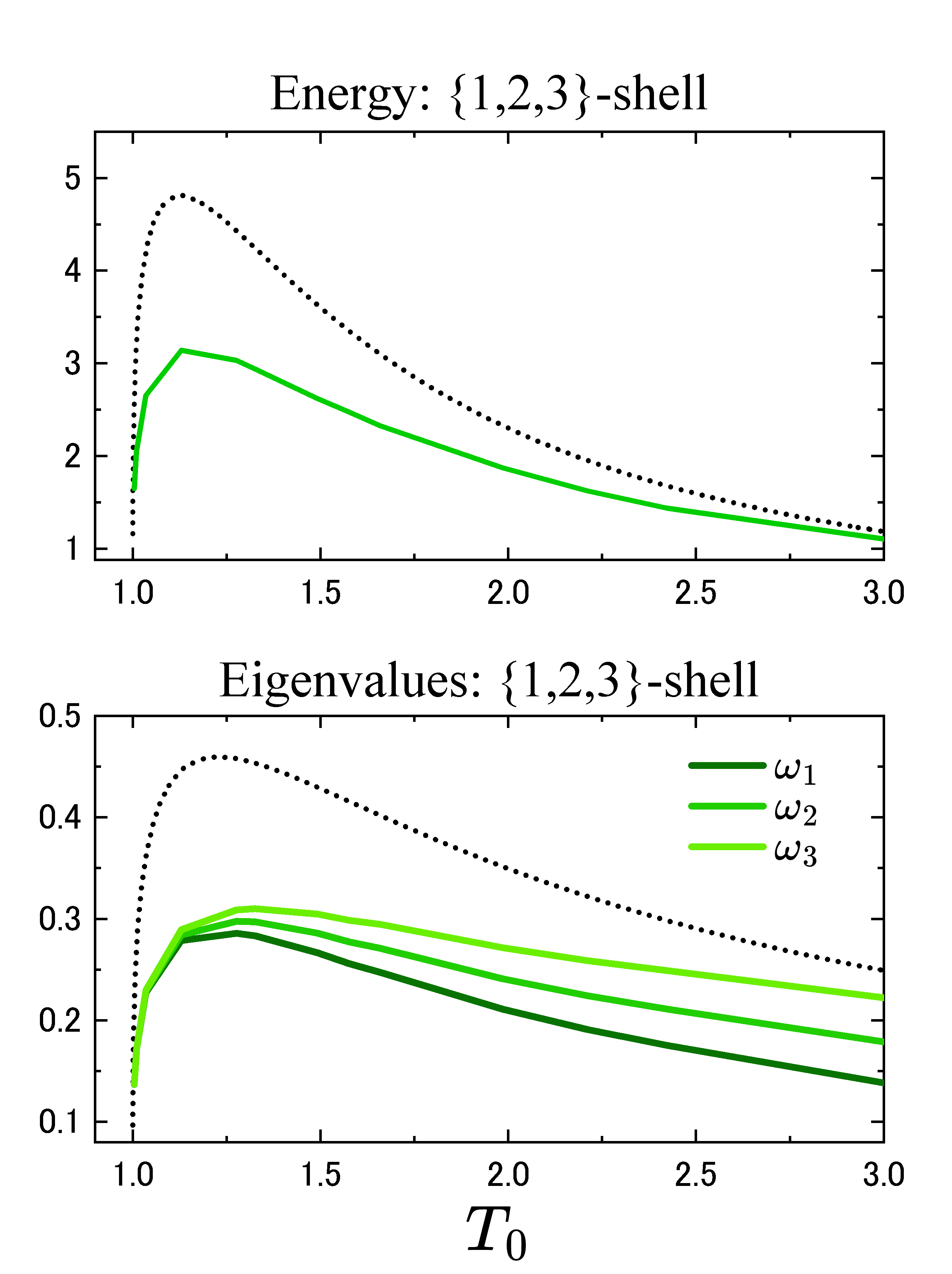}\hspace{-0.1cm}
	\includegraphics[width=0.3\linewidth]{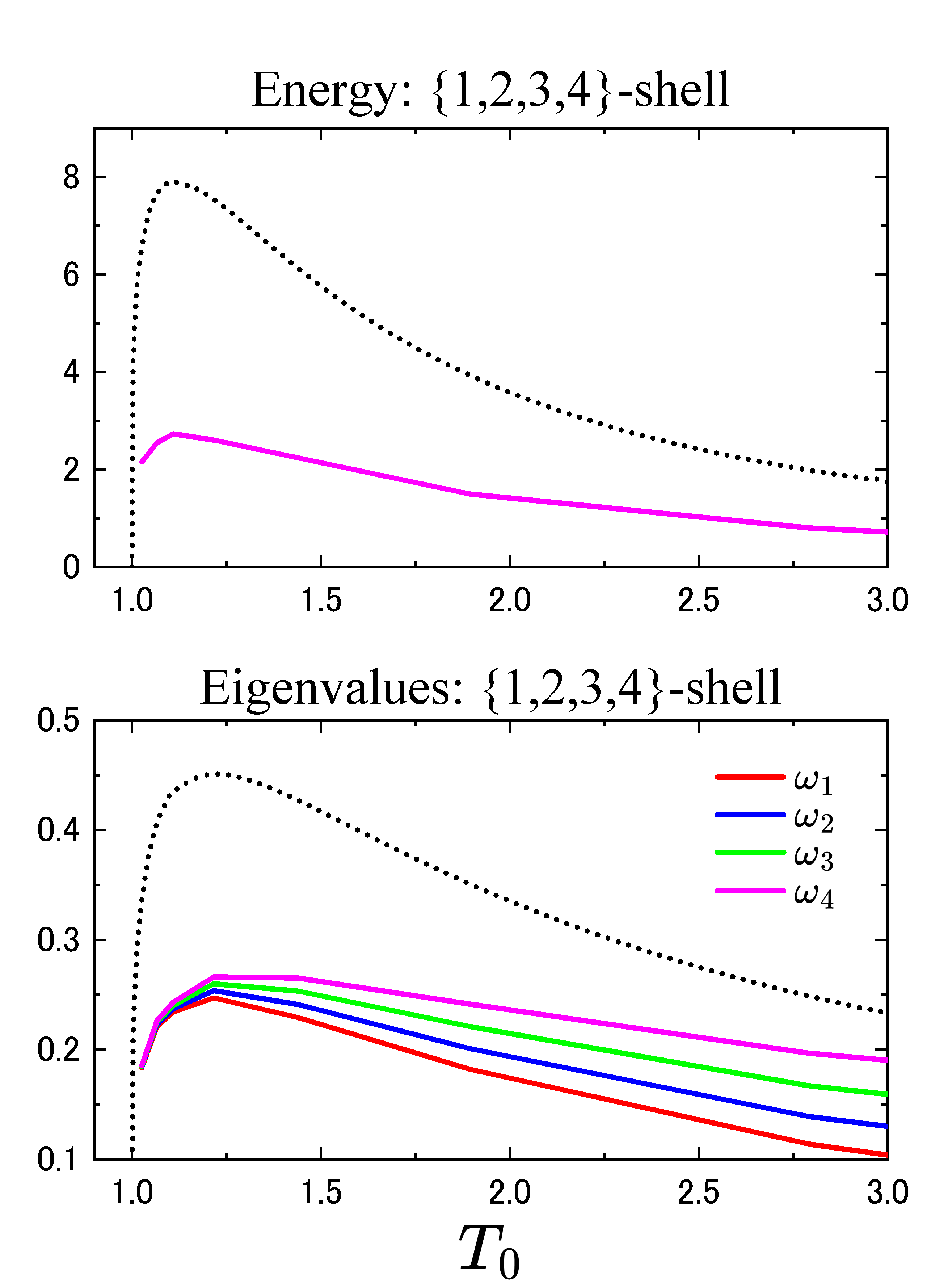}
	\\
	\caption{\label{energy_low}Energies $E$ (upper three panels) 
	and eigenvalues of the Dirac equation $\{\omega_n\}$ (lower three panels)
	in the $\{1,2\}$two-shell (right) with $N_\textrm{f}=6$, 
	$\{1,2,3\}$three-shell (middle) with $N_\textrm{f}=12$, 
	and $\{1,2,3,4\}$-shell with $N_\textrm{f}=20$ (right). 
	The dotted line plots the corresponding approximations for single shells with 
	same fermion numbers. }
\end{figure*}

\section{\label{sec:3}~Numerical method}

To numerically solve the differential equations \eqref{Diraca}--\eqref{EinsteinT}, 
we must impose boundary conditions. As the fermions must be regular at the origin and at infinity,  
we set
\begin{align} 
&\alpha_n|_{r=0}=\beta_n|_{r=0}=0,
\\
&\alpha_n|_{r\to\infty}=\beta_n|_{r\to\infty}=0\,.
\label{spinasymBC}
\end{align}
As localized fermions, require smooth connection to the metric to the standard Schwarzschild metric, 
the fields $T(r),A(r)$ should satisfy
\begin{align}
 T(r)|_{r\to\infty}=A(r)|_{r\to\infty}=1\,.
 \label{metasymBC}
 \end{align}
The standard Frobenius method gives the following asymptotic solutions at the origin:
\begin{align}
\alpha_n(r)&=\alpha_{n0}r^{n}+\ldots,\\
\beta_n(r)&=\frac{1}{2n+1}(\omega_nT_0-m)r^{n+1}+\ldots,\\
T(r)&=T_0-\frac{2G T^2_0}{3}\alpha_{10}^2(2\omega_1T_0-m)r^{2}+\ldots,\\
A(r)&=1-\frac{8G\omega_nT_0^2}{3}\alpha^2_{10}r^{2}+\ldots.~
\end{align}
Here, the new parameter $T_0$ is called the redshift parameter. 
The normalization condition of the spinor is given by
\begin{align} 
&\int\frac{|\Psi_{j_n,k_n}|^2}{2n}\sqrt{\mathrm{-det}(g_{ij})}d^3x
\nonumber \\
&=\int_0^\infty\Bigl(\alpha_n(r)^2+\beta_n(r)^2\Bigr)\frac{T(r)}{\sqrt{A(r)}}dr=1.
\label{spinorNC}
\end{align}
As the normalization condition \eqref{spinorNC} 
and infinite-boundary conditions~\eqref{spinasymBC} and \eqref{metasymBC} 
are difficult to implement numerically, 
we again adopt the innovative approach of Finster~\cite{Finster:1998jqw} and
temporarily impose the weaker conditions
\begin{align}
&\lim_{r\to\infty}T(r)<\infty,~~
\nonumber \\
&\int_0^\infty \Bigl(\alpha_n(r)^2+\beta_n(r)^2\Bigr)\frac{T(r)}{\sqrt{A(r)}}dr=\lambda_n^2<\infty,
\\
&\hspace{5cm}n=1,\cdots,N\,.
\nonumber 
\end{align}
introducing the parameters $\lambda$ and $\tau$ respectively defined as
\begin{align}
&\lambda:=\lambda_1=\lambda_2=\cdots =\lambda_N,~~
\tau=\lim_{r\to\infty}T(r)\,,
\label{BC}
\end{align}
we define the following new functions
\begin{align}
&\tilde{\alpha}_n(\tilde{r})=\sqrt{\frac\tau\lambda}\alpha_n(r),~~
\tilde{\beta}_n(\tilde{r})=\sqrt{\frac\tau\lambda}\beta_n(r)\,
\nonumber \\
&\tilde{A}(\tilde{r})=A(r),~~
\tilde{T}(\tilde{r})=\tau^{-1}T(r),
\nonumber \\
&\tilde{r}:=\frac{r}{\lambda}\,.
\end{align}
These function satisfy \eqref{Diraca}-\eqref{EinsteinT} with new 
parameters $\tilde{m}$ and $\tilde{\omega}_n$, respectively defined as
\begin{align}
&\tilde{m}=\lambda m,~~
\tilde{\omega}_n=\lambda\tau\omega_n\,.
\end{align}

Equations~\eqref{Diraca}-\eqref{EinsteinT} are integrated as follows:\\
(i)~Fixing $m$ and $\omega_1$ and setting appropriate $\omega_2$ and $\omega_3$, 
solve the equations with the shooting parameters $\{\alpha_{n0}\},n=1,\cdots,N$.\\
(ii)~For nodeless solutions, explore some large critical points $r\equiv r_\textrm{crit}$
at which $\alpha_n,\beta_n<0$. \\
(iii)~To smoothly connect the solution to the Schwarzschild solution $A_\textrm{Sch}$, 
define the Wronskian as
\begin{align}
W:=A(r_\textrm{crit})A_\textrm{Sch}'(r_\textrm{crit})-A'(r_\textrm{crit})A_\textrm{Sch}(r_\textrm{crit})\,.
\nonumber 
\end{align}
(iv)~Under the proper normalization condition \eqref{BC}, define
\begin{align}
\Lambda_n:=\biggl|1-\frac{\lambda_n}{\lambda_1}\biggr|,~~n=1,\cdots,N
\end{align}
To conserve computational resources, we apply a somewhat imprecise convergence criterion at this time: 
that is 
\begin{align}
W<10^{-6},~~\Lambda_n<10^{-3}\,.
\end{align}
Gradually changing $\omega_2$ and $\omega_3$, repeat steps (ii)-(iv) until the solution converges.
To reduce the computational cost, we adopt a simplified algorithm for the four-shell analysis. 
We first investigate rapidly converging solutions $\alpha_1,\beta_1$ in step (i). 
When the solutions are sufficiently small at $r=r_0$, they are set to zero at $r>r_0$.
The remaining components $\alpha_n,\beta_n,~~n > 1$ are then obtained.  

Although this method is applicable to any number of shells, precise calculations (especially 
of the shooting parameters in (i)) are required for large shell numbers.  
Even the three-shell calculation require a quadruple-precision floating-point
format (binary128) to reduce the accumulation of numerical errors. 
This paper analyzes up to the four shells; an analysis of five shells requires a more sophisticated numerical 
algorithm.

\section{\label{sec:4}~Numerical results}

Leith \textit{et al}.~\cite{Leith:2020jqw} simply assumed that every fermion is grouped in a distinct shell with 
a given angular momentum. 
All characteristic behaviors emerging from the extension to many fermions can be 
attributed to the strong gravity imposed by the large number of particles and 
the substantial angular momentum of each particle. 
Therefore, the ground states can be identified in few-particle cases.

\subsection{Model solutions}

We first obtained the $\{3\}$-, $\{6\}$-, and $\{10\}$-shell solutions with   
fermion numbers of $N_\textrm{f}=6,12,20$, respectively, in the single-shell approximation.   
Using~\eqref{Ycomplete} under the normalization condition, the
number density of the spinors is given by
\begin{align}
d_n(r):=2n\Bigl(\alpha_n(r)^2+\beta_n(r)^2\Bigr)\frac{T(r)}{\sqrt{A(r)}}
\end{align}
where the integration estimates the fermion number $2n$ in the $n$th shell  
\begin{align}
\int_0^\infty d_n(r)dr = 2n\,.
\end{align}
Figure~\ref{1shell} plots the metric functions $A(r)$ and $T(r)$, 
the spinor components $\alpha(r),\beta(r)$ and the 
number densities $d_n(r)$ 
for $T_0=2.0$ and $60.0$. 
These results correspond to those in ~\cite{Finster:1998jqw,Leith:2020jqw}
~\footnote{Minor differences between our results and those in~\cite{Leith:2020jqw} 
are caused by the normalization condition of the spherical harmonics.}.  

Figure~\ref{234shell} presents the results of our multishell model. 
Here we obtained the $\{1,2\}$-, $\{1,2,3\}$-,$\{1,2,3,4\}$-shell solutions 
for fermion numbers of $N_{\textrm f}=6,12,20$, respectively, 
at low- and high red-shifts.   
As $T_0$ increases, the $A(r)$ and $T(r)$ plots reveal increasing compactifications of  
the objects and the number density plot apparently broaden, particularly in the outer shell. 
The peaks of the spinor components begin delocalizing as similarly observed in 
the single-shell approximation. 
The diminished peak in the outer shell, despite possessing the
number of fermions, 
is worth comparing with the corresponding shell solution in Fig.\ref{1shell}. 
In the low-redshift case, the single-shell solution with $T_0=2.0$ is located around 
$r: (3.0, 11.0)$, whereas the two-shell solution with $T_0=2.5$ appears 
about $r: (0.2, 12.0)$.  
In the high-redshift case, the single shell solution and 
the two-shell solution with $T_0=60.0$ are distributed around $r: (0.2, 30.0)$ and $r: (0.02, 25.0)$, 
respectively, clarifying more internalized and more compact 
solutions in the two-shell model. 
In Table \ref{tab:example}, we summarize the comparison of the radial distributions of the number density 
between the single-shell and the several multishell solutions corresponding to Figs.\ref{1shell} and \ref{234shell}.
Such behavior will likely to become more obvious when calculating the averaged density of the objects.   
The averaged density is discussed in the next section.

\begin{figure}[htbp]
	\centering 
	\includegraphics[width=0.8\linewidth]{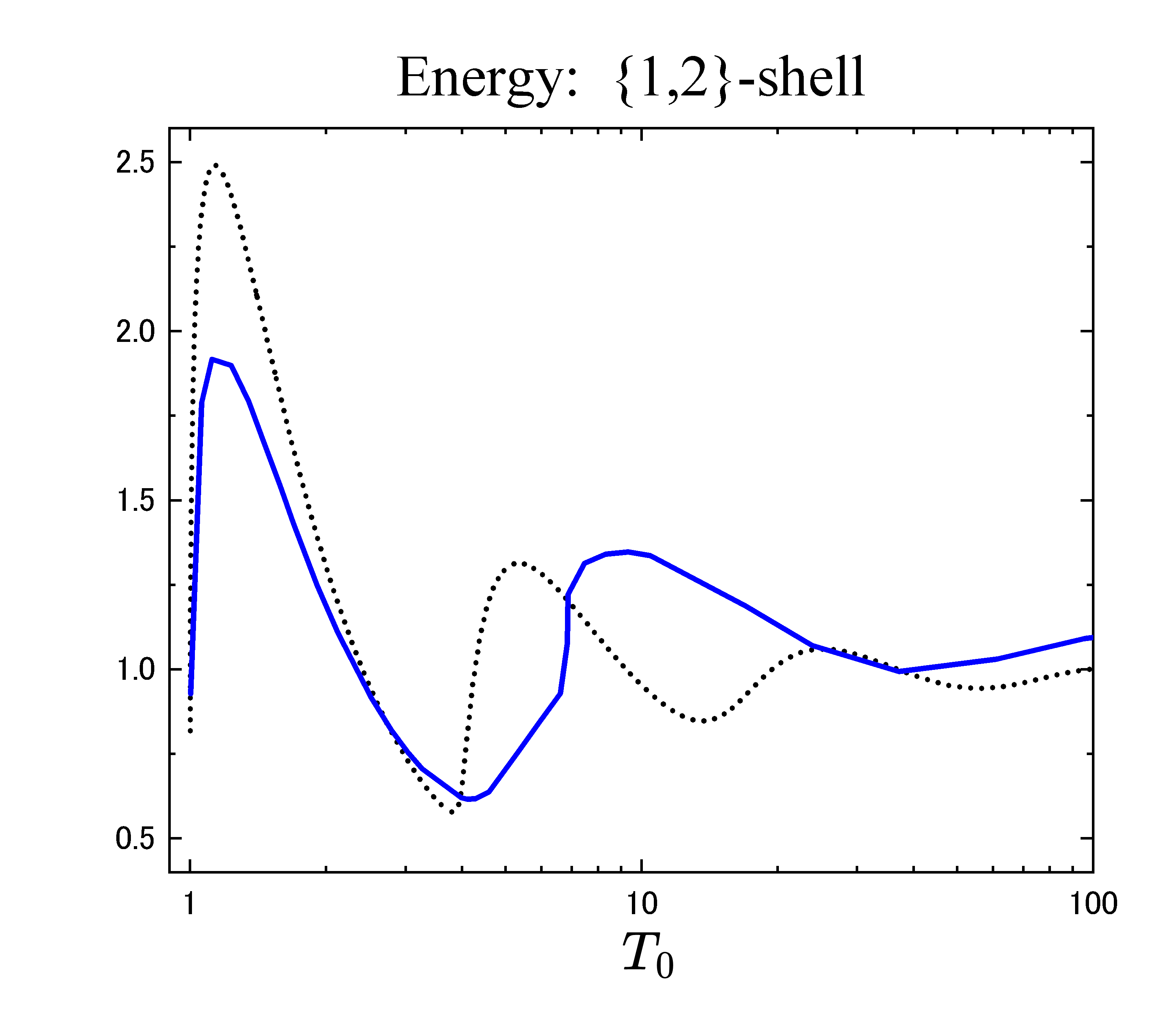}
	\includegraphics[width=0.8\linewidth]{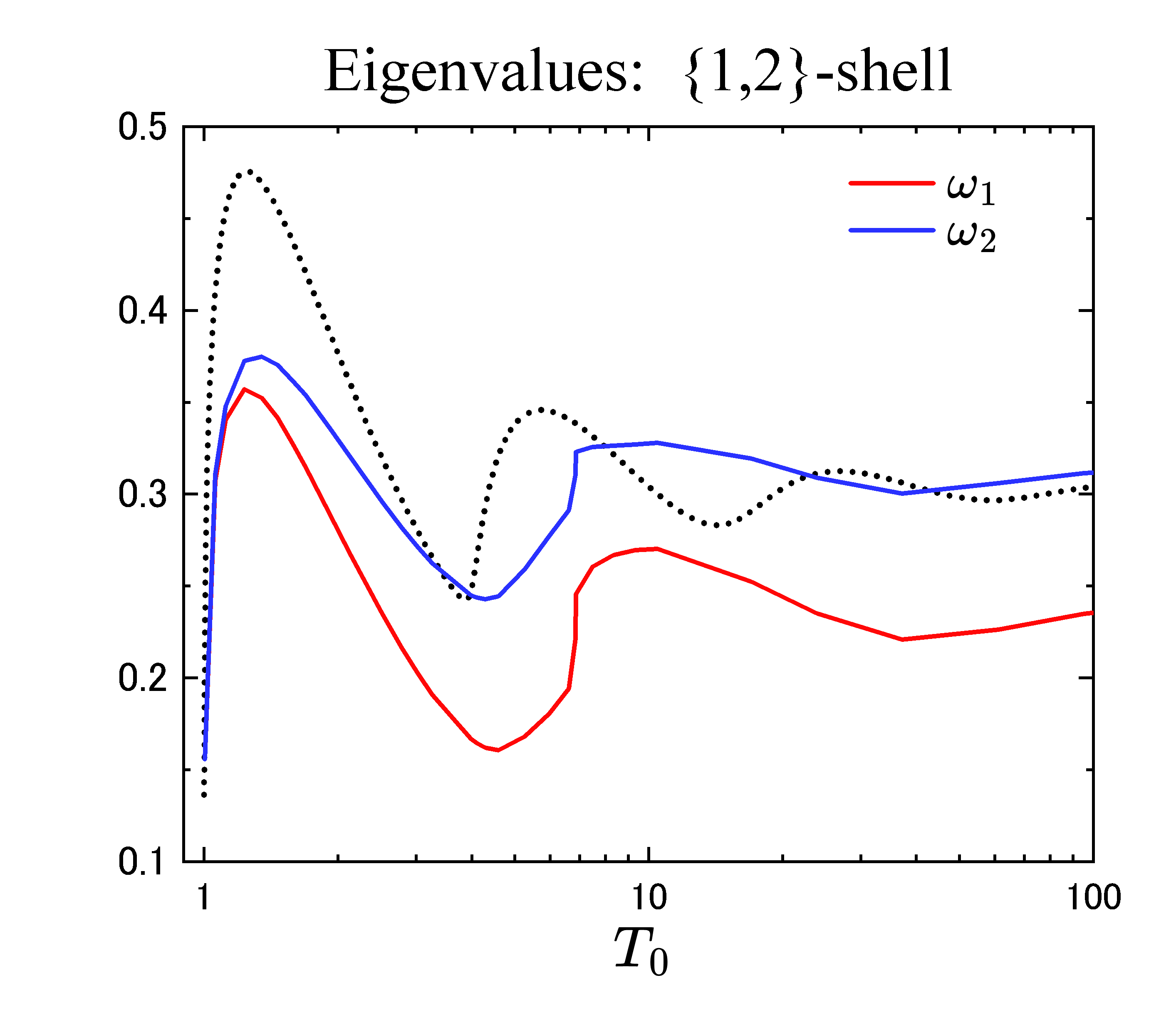}
	\caption{\label{energy2}Energy $E$ and eigenvalues $\omega_n,n=1,2$ 
	as functions of $T_0$ in the $\{1,2\}$two-shell model. 
	The dashed lines are the corresponding results of the single-shell approximation. }
\end{figure}

\begin{figure}[htbp]
\centering 
	\includegraphics[width=0.8\linewidth]{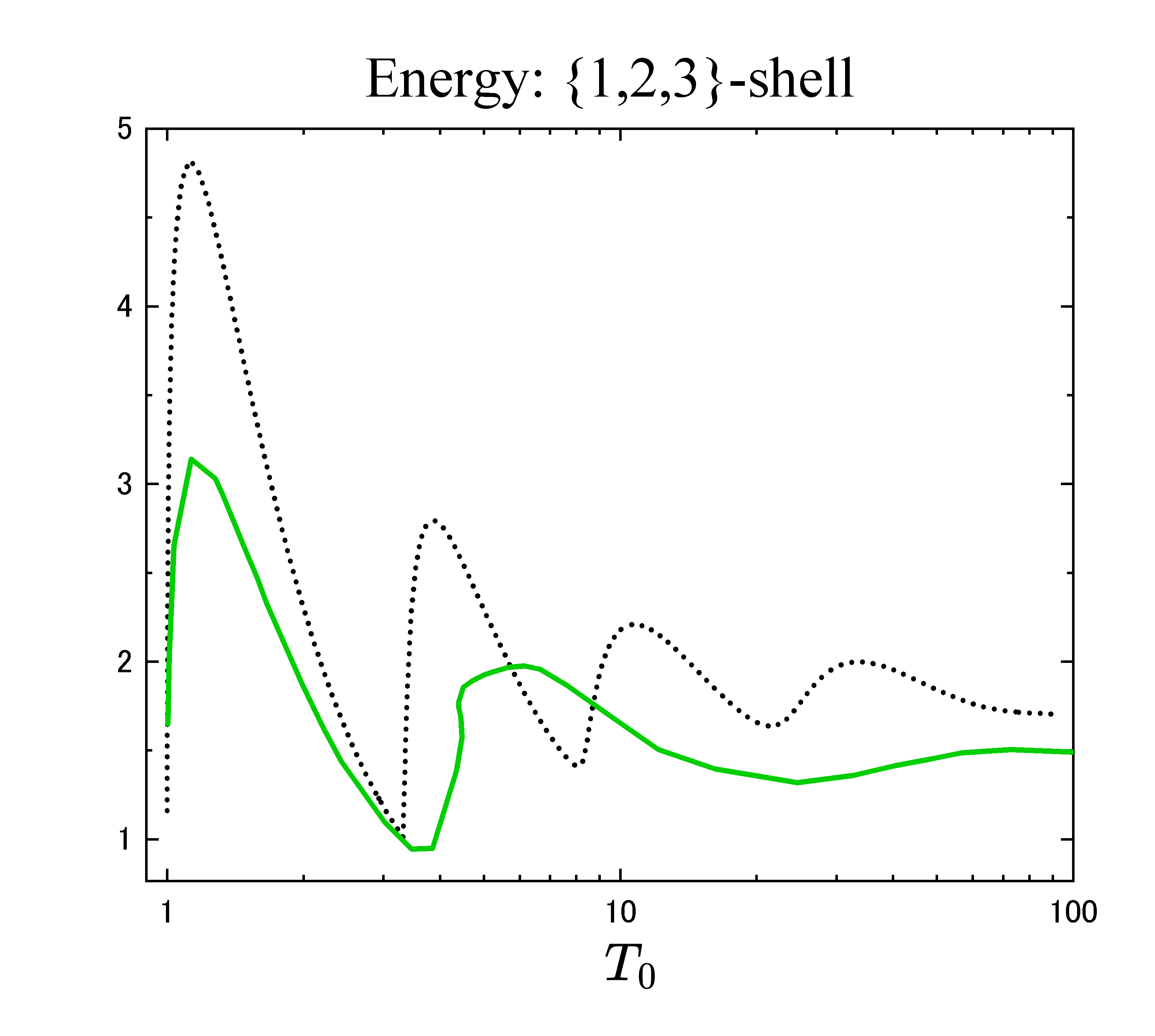}
	\includegraphics[width=0.8\linewidth]{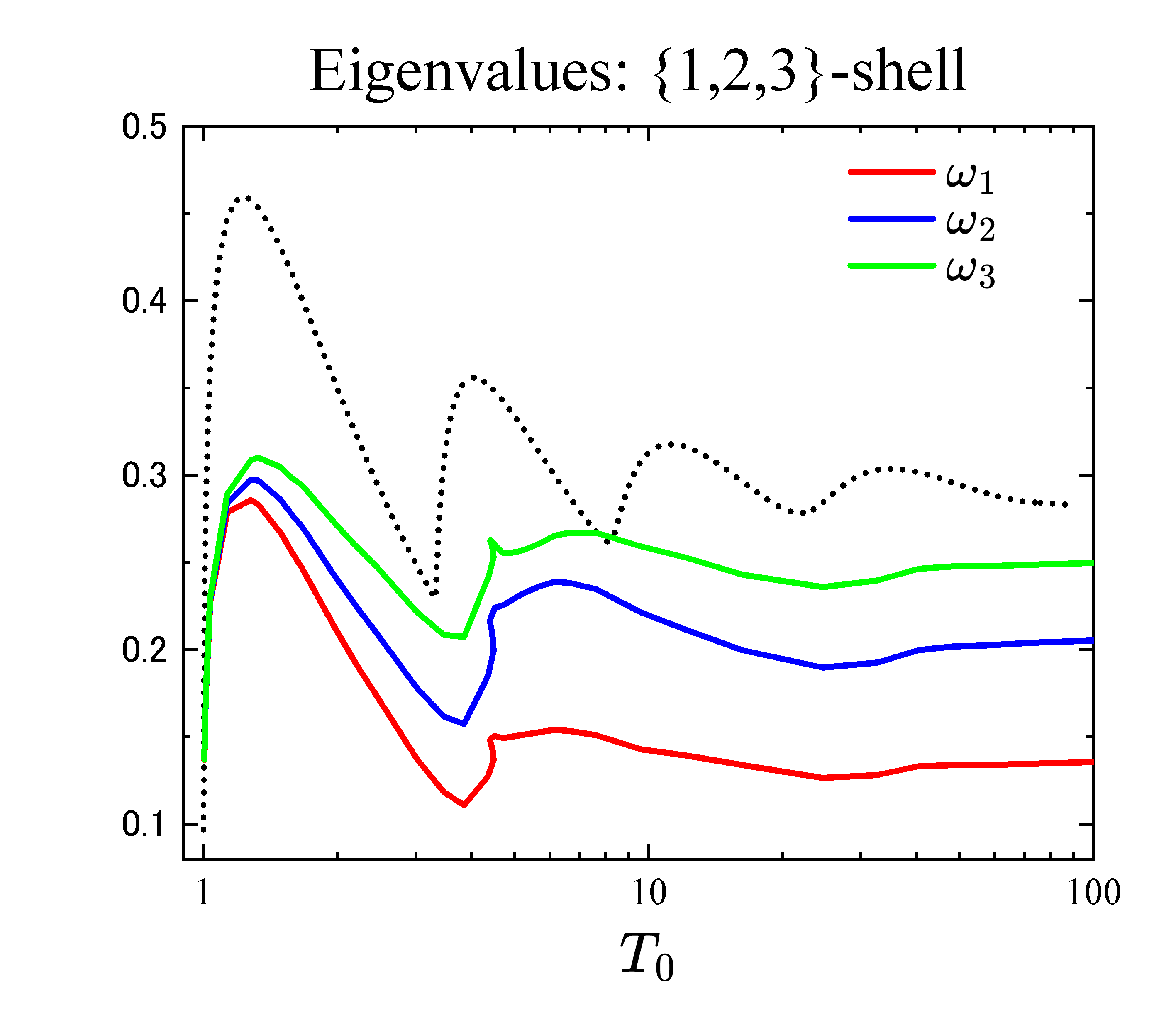}
	\caption{\label{energy3}
	As for Fig.\ref{energy2}, but showing the results of the $\{1,2,3,\}$three-shell solution with	
	eigenvalues $\omega_n,~~n=1,2,3$.}
\end{figure}

\begin{figure*}[t]
	\centering 
	\includegraphics[width=0.33\linewidth]{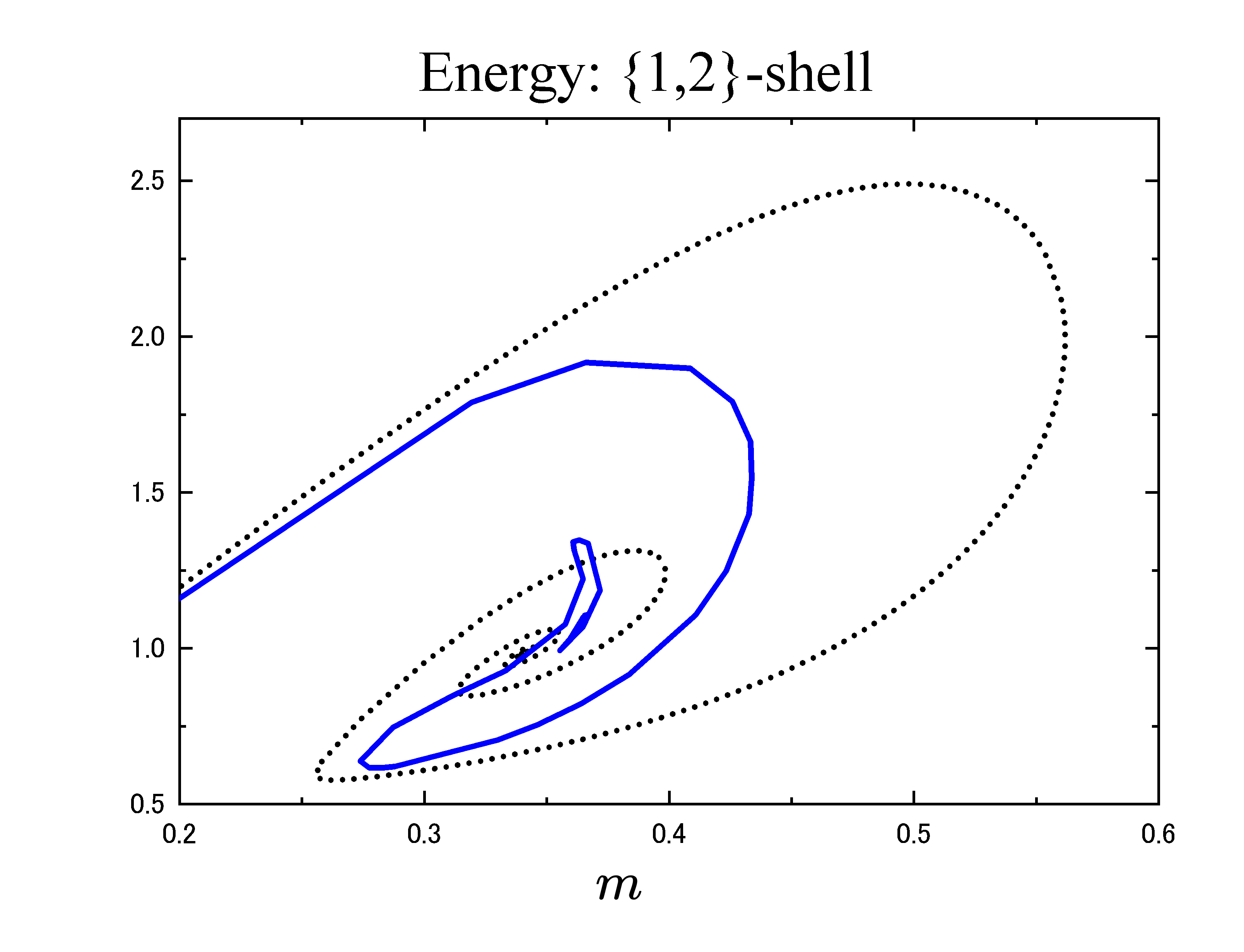}\hspace{-0.4cm}
	\includegraphics[width=0.33\linewidth]{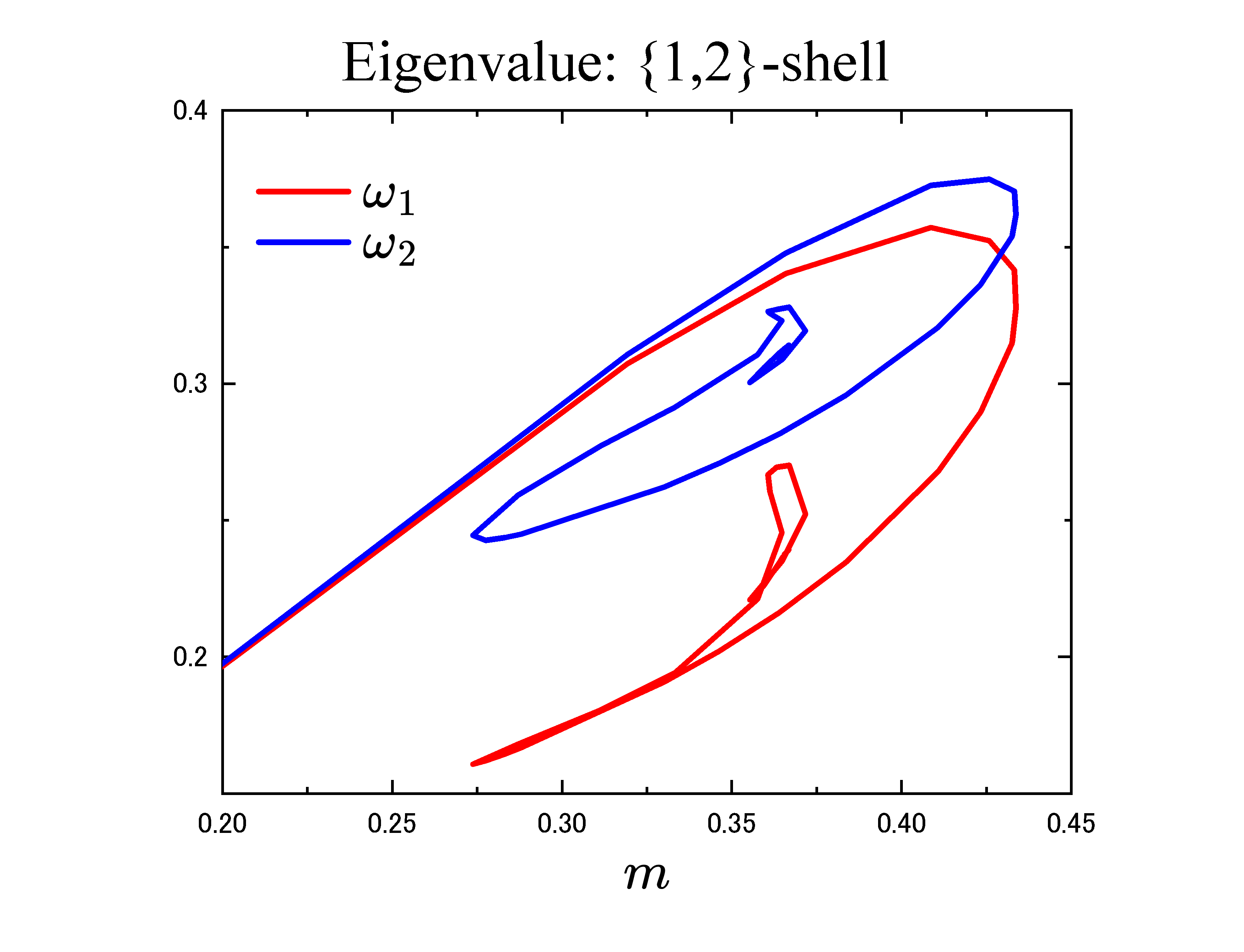}\hspace{-0.6cm}
	\includegraphics[width=0.33\linewidth]{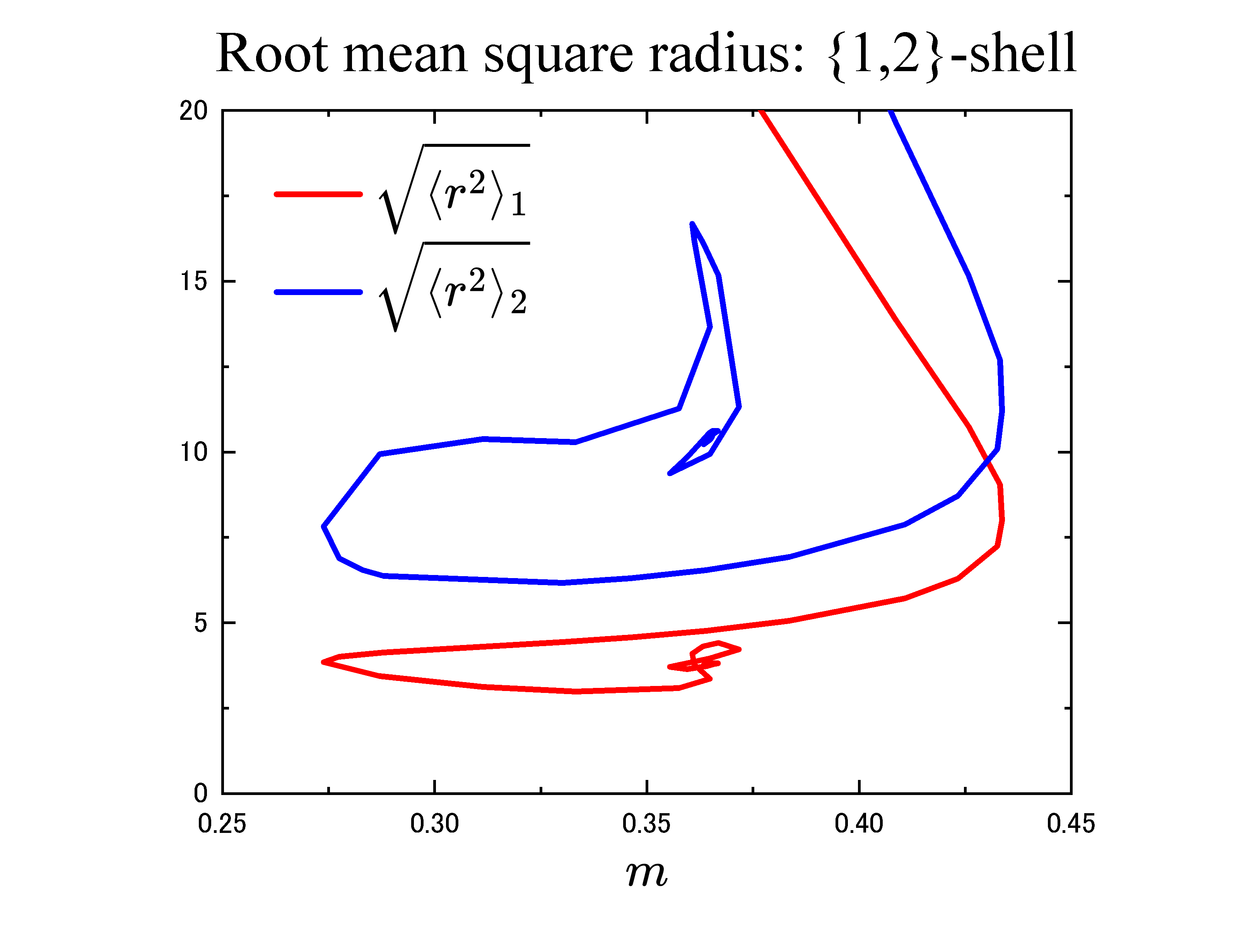}
	\caption{\label{2shell_phase}Phase diagrams of mass-energy $m$--$E$ (left), 
	mass-eigenvalues $m$--$\omega_n$ (middle), 
	and mass-radius $m$--$\sqrt{\langle r^2\rangle_n}$, $n=1,2$ (right),  
	of the $\{1,2\}$two-shell solution. In the mass-energy diagram, the dotted curve plots 
	the corresponding single-shell
	approximations with the same fermion numbers $N_\textrm{f}=6$.}
\end{figure*}

\begin{figure*}[t]
	\centering 
	\includegraphics[width=0.33\linewidth]{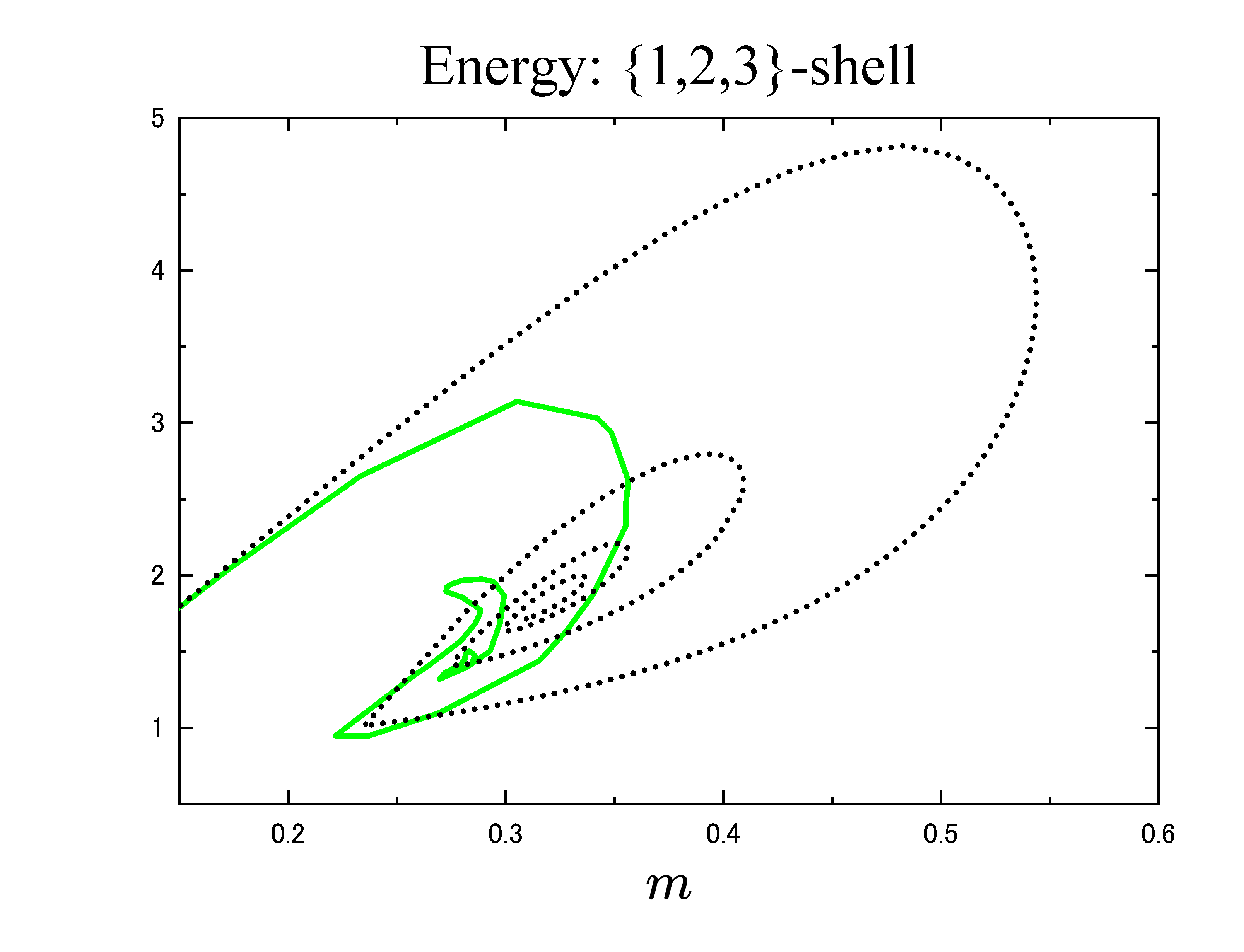}\hspace{-0.4cm}
	\includegraphics[width=0.33\linewidth]{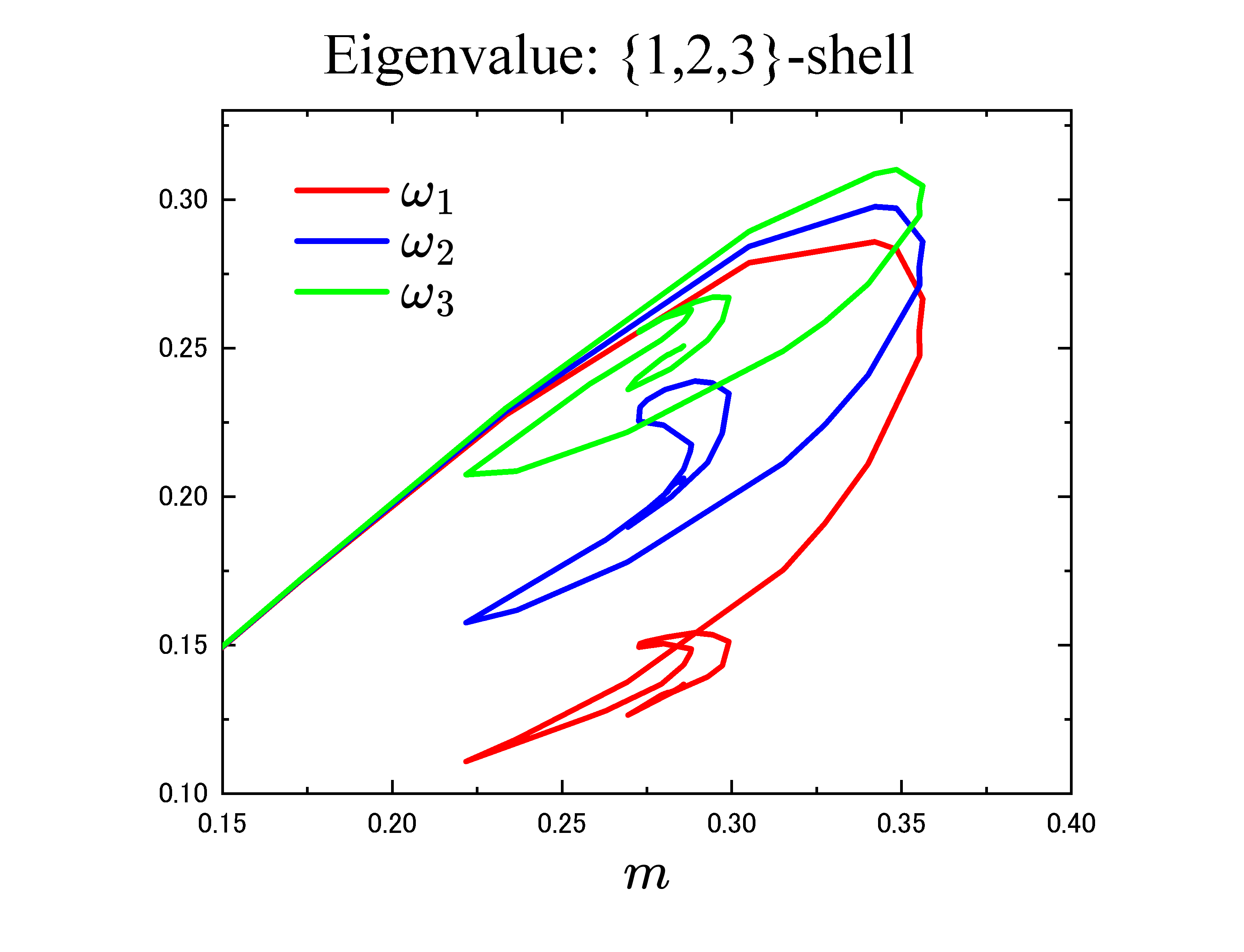}\hspace{-0.6cm}
	\includegraphics[width=0.33\linewidth]{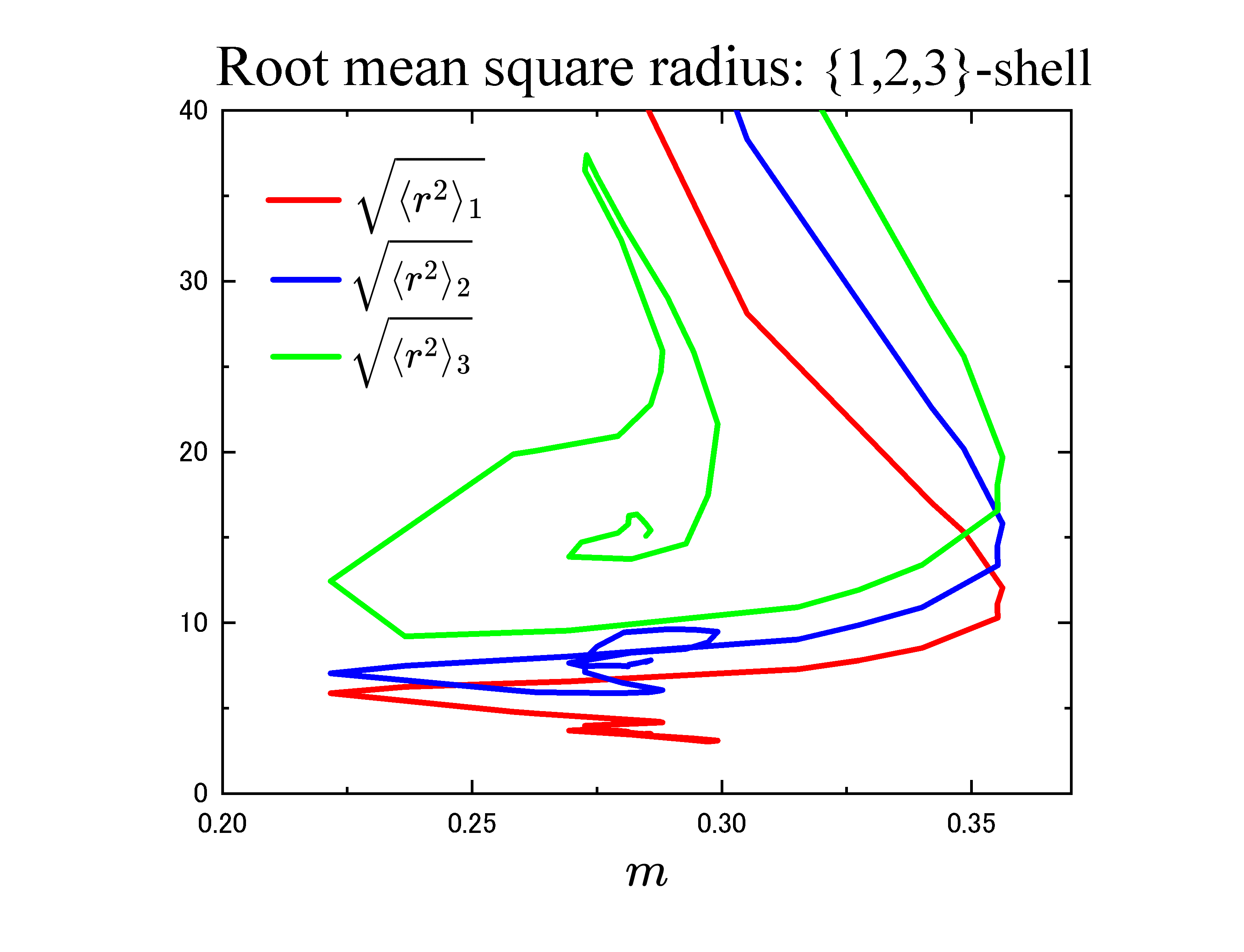}
	\caption{\label{3shell_phase}As for Fig.\ref{2shell_phase}, but for the $\{1,2,3\}$three-shell solution
	with root mean square radii $\sqrt{\langle r^2\rangle_n}$, $n=1,2,3$.
	In the mass-energy diagram, the dotted curve plots the corresponding single-shell
	approximations with the same fermion number $N_\textrm{f}=12$.}
\end{figure*}

\begin{figure*}[t]
	\centering 
	\includegraphics[width=0.4\linewidth]{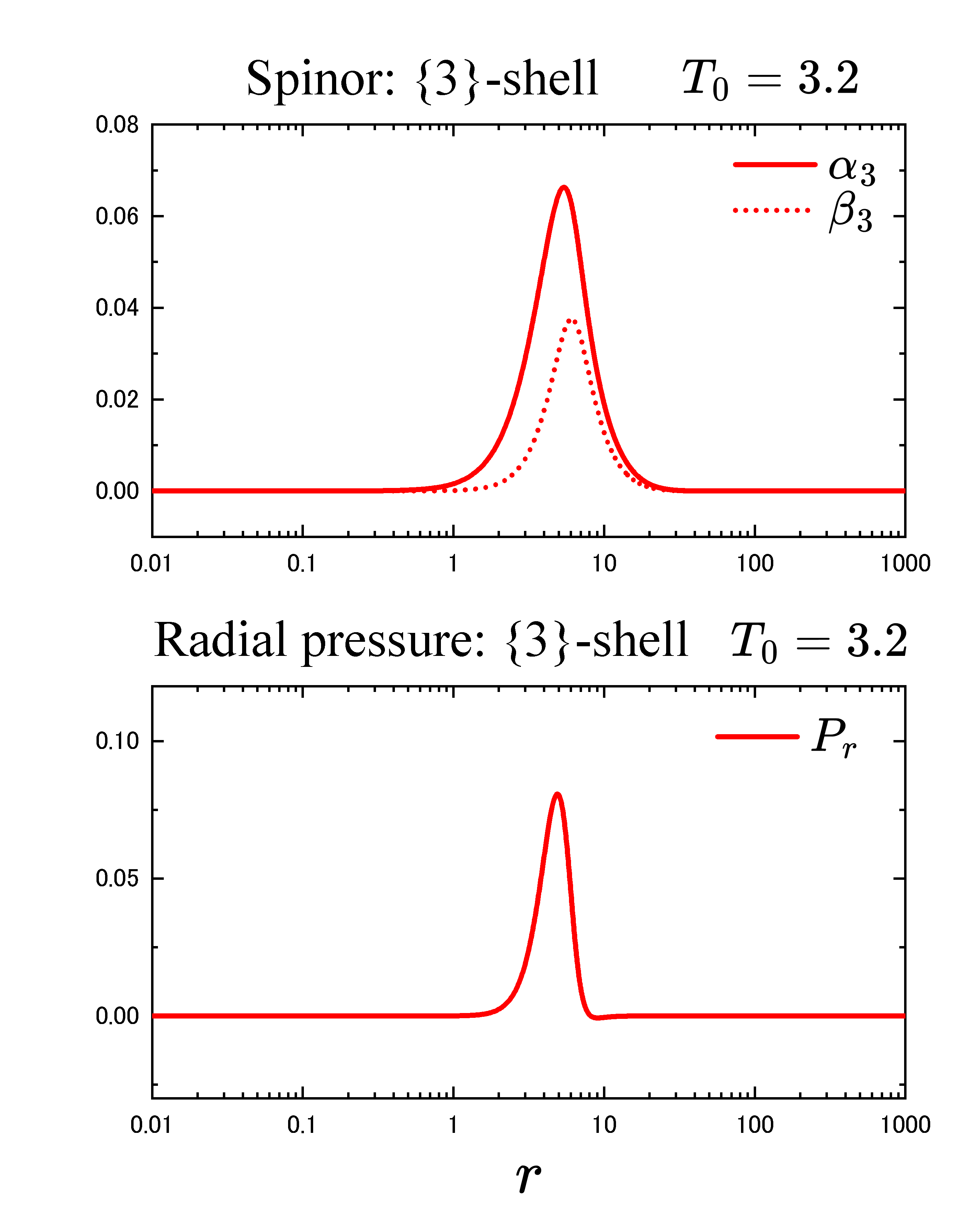}\hspace{-0.3cm}
	\includegraphics[width=0.4\linewidth]{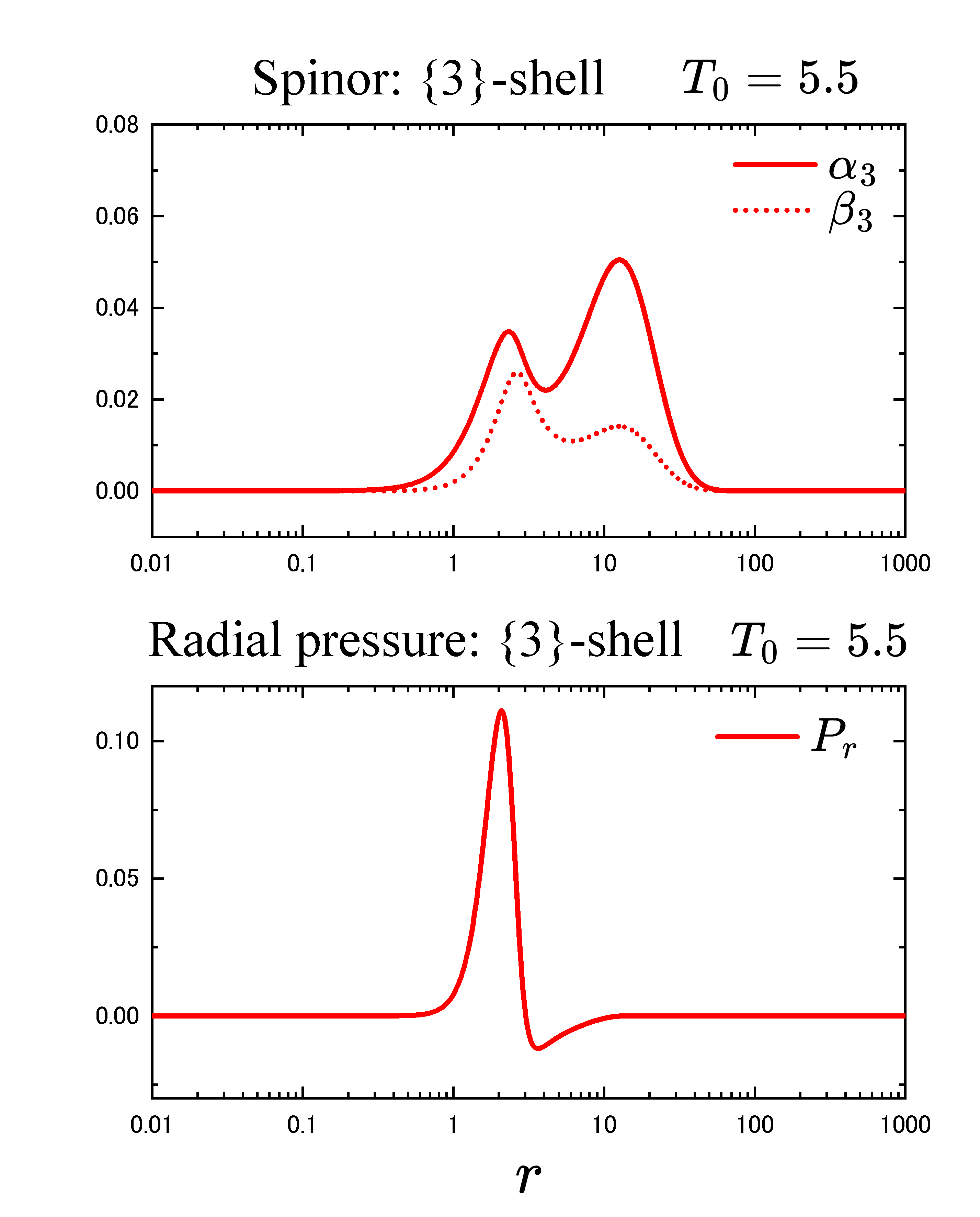}
	\caption{\label{1shell_NP}Spinor (top) and radial-pressure (bottom) solutions
	of the $\{3\}$single-shell model with $T_0=3.2, 5.5$.
	A negative-pressure region emerges when the shell starts to fractionize ($T_0=5.5$)
	as follows: 
 	similarly discussed in~\cite{Andreasson:2025uir}.}
\end{figure*}

\begin{figure*}[htbp]
	\centering 
	\includegraphics[width=1\linewidth]{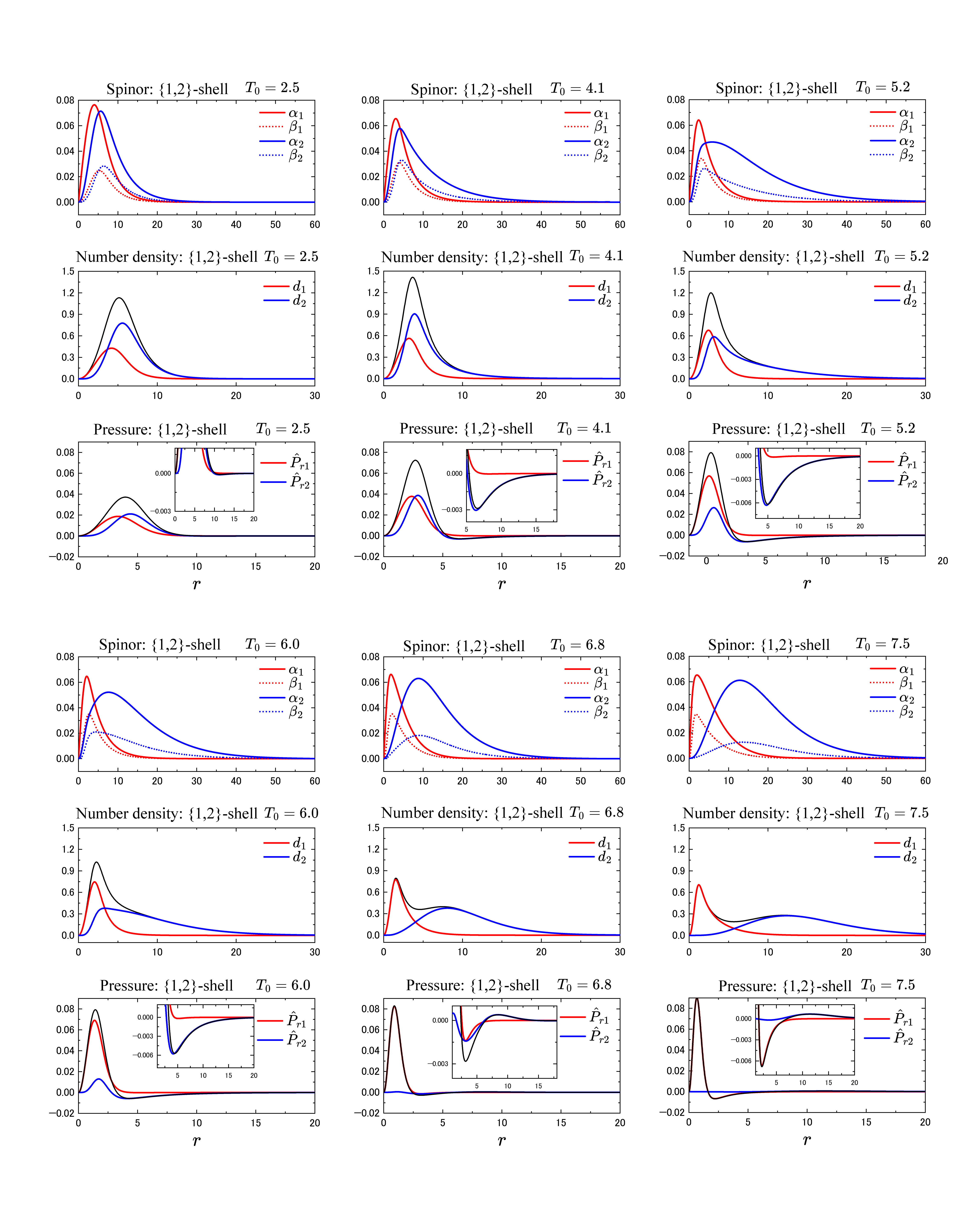}
	\caption{\label{2shell_NP}Spinor (top), number density (middle) and     
      radial-pressure (bottom)
	solutions of the $\{1,2\}$two-shell model with $T_0=2,5, 4.1, 5.2, 6.0, 6.8, 7.5$. 
	The insets in the radial-pressure plots are enlargements to clarify the negative pressure area. 
    The total values of the pressure and the number density as a sum of each shell are shown by the black line.}
\end{figure*}

\begin{figure*}[htbp]
	\centering 
	\includegraphics[width=1\linewidth]{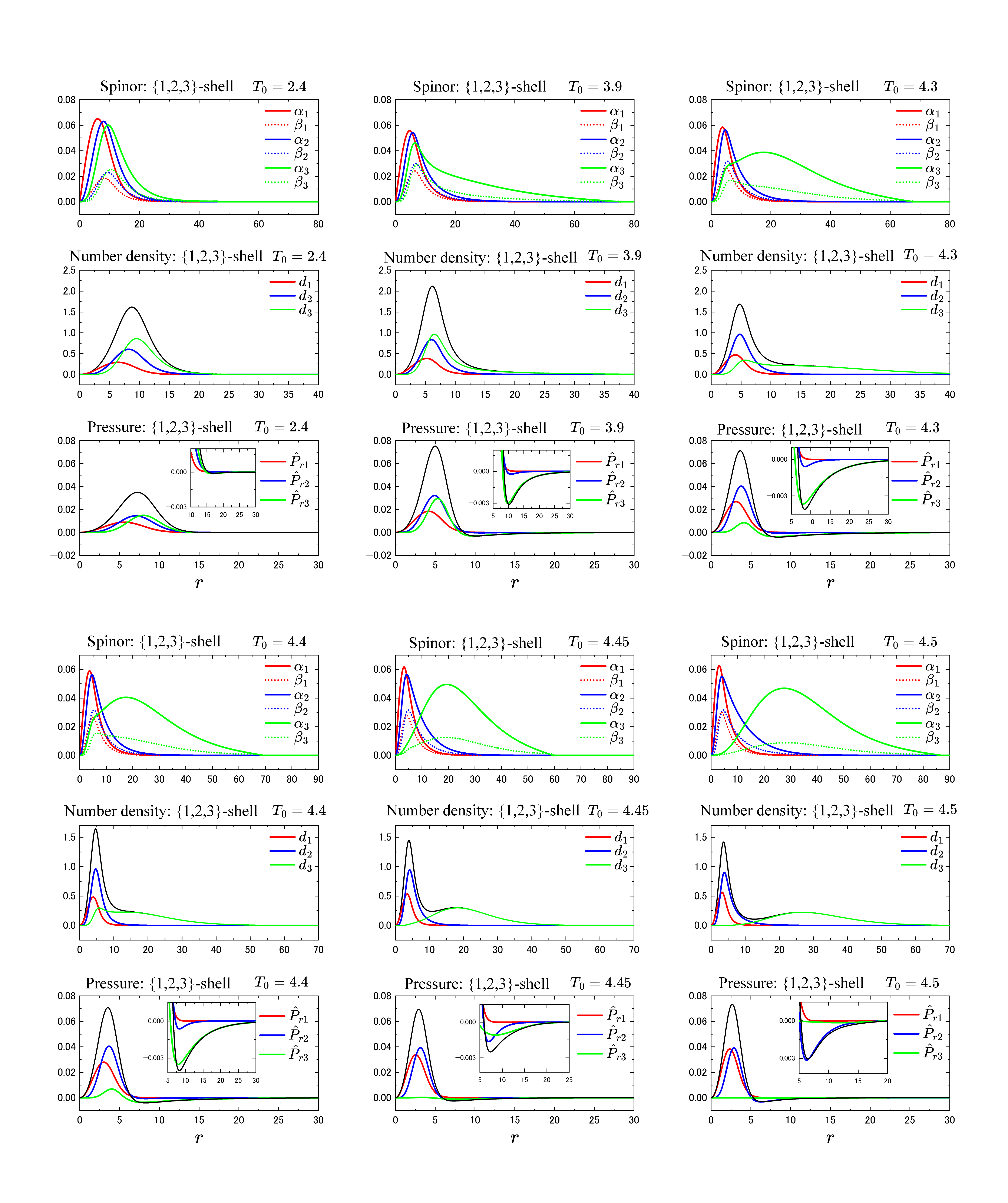}
	\caption{\label{3shell_NP}Spinor, number density and radial-pressure solutions 
	of the $\{1,2,3\}$three-shell model with $T_0=2,4, 3.9, 4.3, 4.4, 4.45, 4.5$. 
	The insets in the radial-pressure plots are enlargements to clarify the negative-pressure area. 
	The total values of the pressure and the number density as a sum of each shell are shown by the black line.}
\end{figure*}

\subsection{Energies and eigenstates}

This subsection investigates two energy quantities of the fermions at  different redshifts $T_0$. 
First, we computed the matter energy by integrating the energy-momentum tensor $T_{tt}$ \eqref{EMt}
\begin{align}
E:=&\int^\infty_0T_{tt}\sqrt{-\textrm{det}(g_{ij})}d^3x\nonumber\\
&=\sum_{n=1}^N2n\omega_n\int^\infty_0\Bigl(\alpha_n(r)^2+\beta_n(r)^2\Bigr)\dfrac{dr}{\sqrt{A}}
\end{align}
Note that in a flat background space-time, this energy reduces to $E=\sum_{n=1}^N 2n\omega_n$. 
Second, we obtain the eigenvalues~$\{\omega_n\}$ of the Dirac equations~\eqref{Diraca},\eqref{Diracb}. 
Figure~\ref{energy_low} plots the energies of the $\{1,2\}$-shell, $\{1,2,3\}$-shell, and $\{1,2,3,4\}$-shell
models at low-redshift $T_0$. 
The results are consistently lower than the corresponding single-shell approximations with the same
fermion numbers ($6,12,20$, respectively).  
The behavior is expected because similarity to the nucleic and atomic models, 
a collection of closed (fully occupied) shells enables the existence of energetically stable states. 
 
However, at higher redshifts $T_0$, the behavior becomes complicated and the model is no longer simply
analogous to the atomic model.  
The following detailed of this phenomenon is simplest examples, namely, the two-shell and three-shell 
solutions. 
Figure~\ref{energy2} presents the energy and eigenvalues at larger $T_0$. 
The energies exhibit typical damped oscillations related to large structural changes such as peak 
fragmentation of the solutions (see the next subsection). 
The first increase in the single-shell energy $E$ at $T_0\sim 4$ is followed by an increase in 
the two-shell energy, which eventually surpasses the single-shell energy increase at $T_0\sim 7$. 
Beyond the critical point, the energy excitation of the two-shell always exceeds that of the single-shell.
The eigenvalues of the two-shell $\omega_1,\omega_2$ behave similarly to the energy solutions. 
In particular, the second eigenvalue $\omega_2$ with angular momentum $j_2=\frac{3}{2}$ 
exceeds that of the single-shell with $j_3=\frac{5}{2}$ after the critical point.
This result deviates from the criterion of the standard shell model indicating that 
gravitation plays a dominant role even in few-fermion cases.    
Figure \ref{energy3} plots the corresponding solutions of the three-shell model. 
In this model, the energy $E_0$ at large $T_0$ is almost 
lower than the energy of the single-shell model. We posit that multishell solutions are a candidate ground states of the system.
 The four-shell solution behaves similarly to the three-shell one, suggesting 
that a large number of fermionic self-gravitating systems will recover the criterion 
determining a nucleus or atomlike property.

\begin{figure*}[htbp]
	\centering 
	\includegraphics[width=0.4\linewidth]{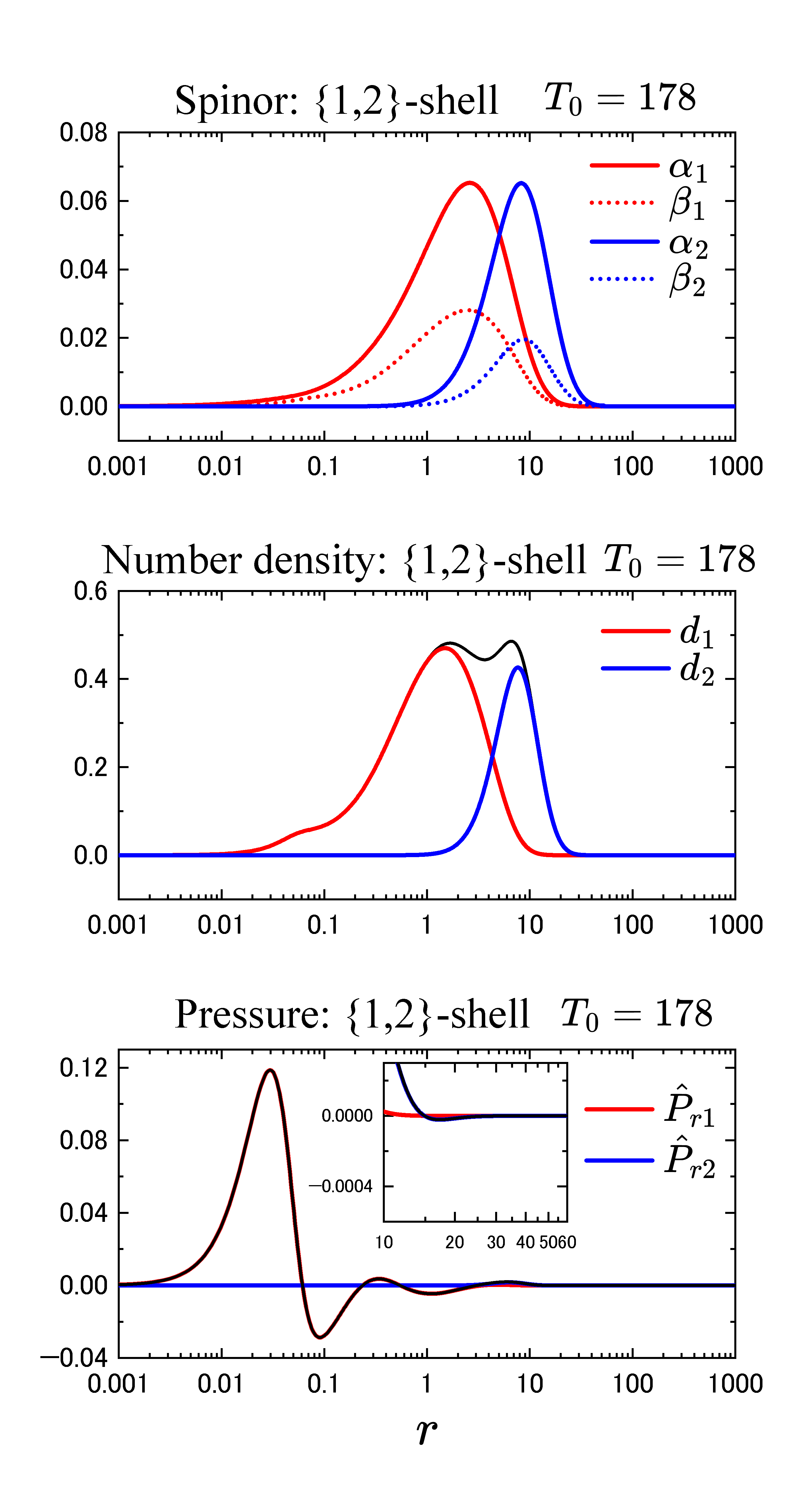}
	\includegraphics[width=0.4\linewidth]{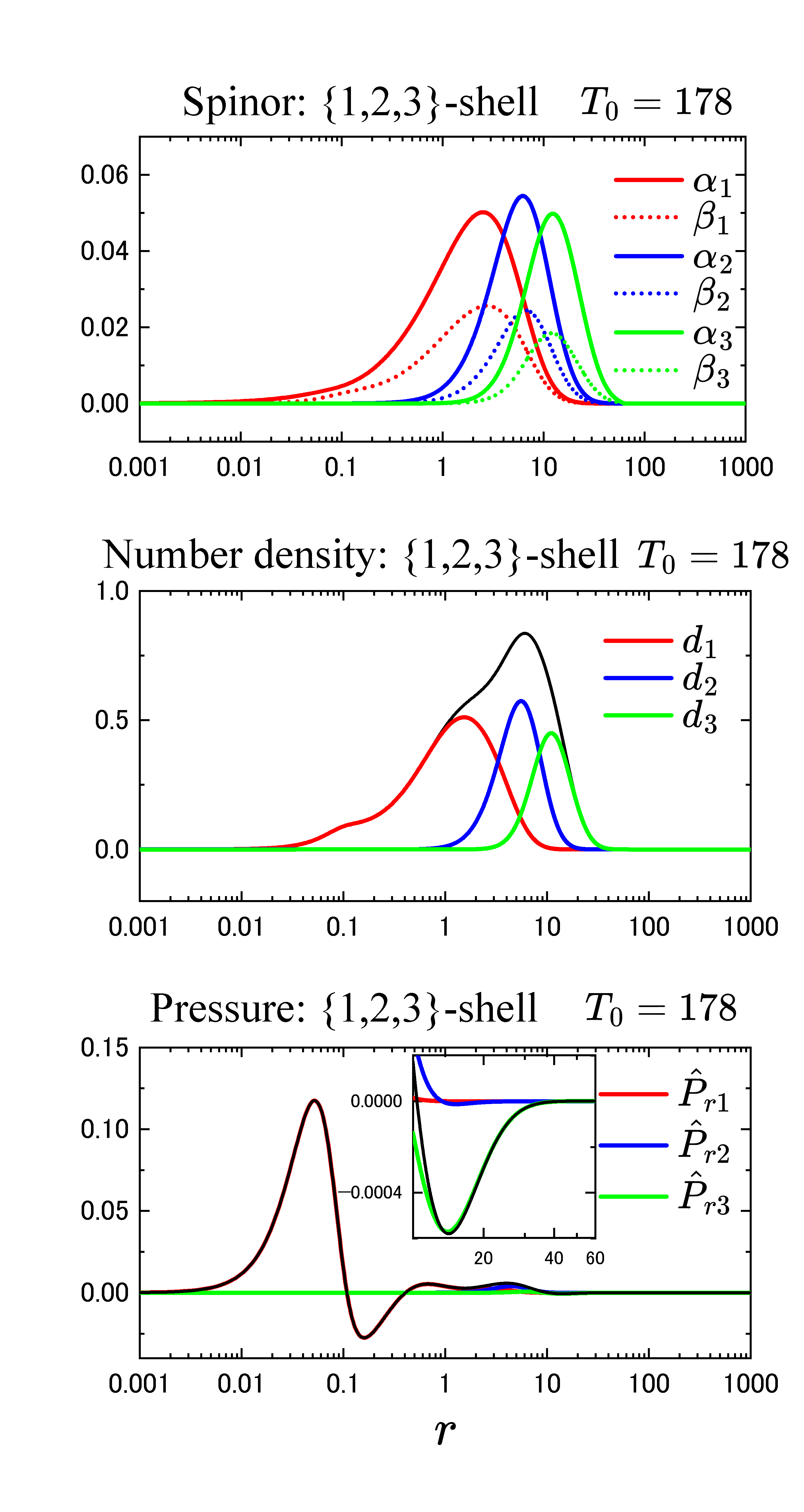}
	\caption{\label{shells_high}
	High-redshift ($T_0=178.0$) solutions of the spinors, number densities and radial-pressure of 
	the $\{1,2\}$two-shell (left) and $\{1,2,3\}$three-shell models (right). 
	The enlarged pressure plots (insets) clarify the 
	negative-pressure region in the farthest outside region.  
	The total values of the pressure and the number density as a sum of each shell are shown by the black line.}
\end{figure*}

\subsection{Phase diagrams}

This subsection displays the phase diagrams obtained by continuously varying the model parameters. 
To clarify the properties of the solutions, we introduce the root mean square radius of the $n$th shell:
\begin{align}
\sqrt{\langle r^2\rangle_n}&:=\biggl[\int r^2|\Psi_{j_n,k_n}|^2\sqrt{-\textrm{det}(g_{ij})}d^3r\biggr]^{\frac{1}{2}}
\nonumber \\
&=\biggl[\int_0^\infty r^2(\alpha_n^2+\beta_n^2)\frac{T(r)}{\sqrt{A}}dr\biggr]^{\frac{1}{2}}\,.
\label{rmsr}
\end{align}
Figure~\ref{2shell_phase} presents the mass-eigenvalues, the mass-energy, and the mass-radius 
phase
diagrams of the $\{1,2\}$-shell solutions at various $T_0$. 
Our model typically presents spiraling curves which are shrunken from that of the corresponding 
single shell approximation. 
The eigenvalues satisfy $\omega_n<m,~~n=1,2$, indicating that the eigenstates are bounded. 
Figure~\ref{3shell_phase} presents the corresponding phase plots of the $\{1,2,3\}$-shell solutions.

\begin{figure*}[t]
	\centering 
	\includegraphics[width=0.48\linewidth]{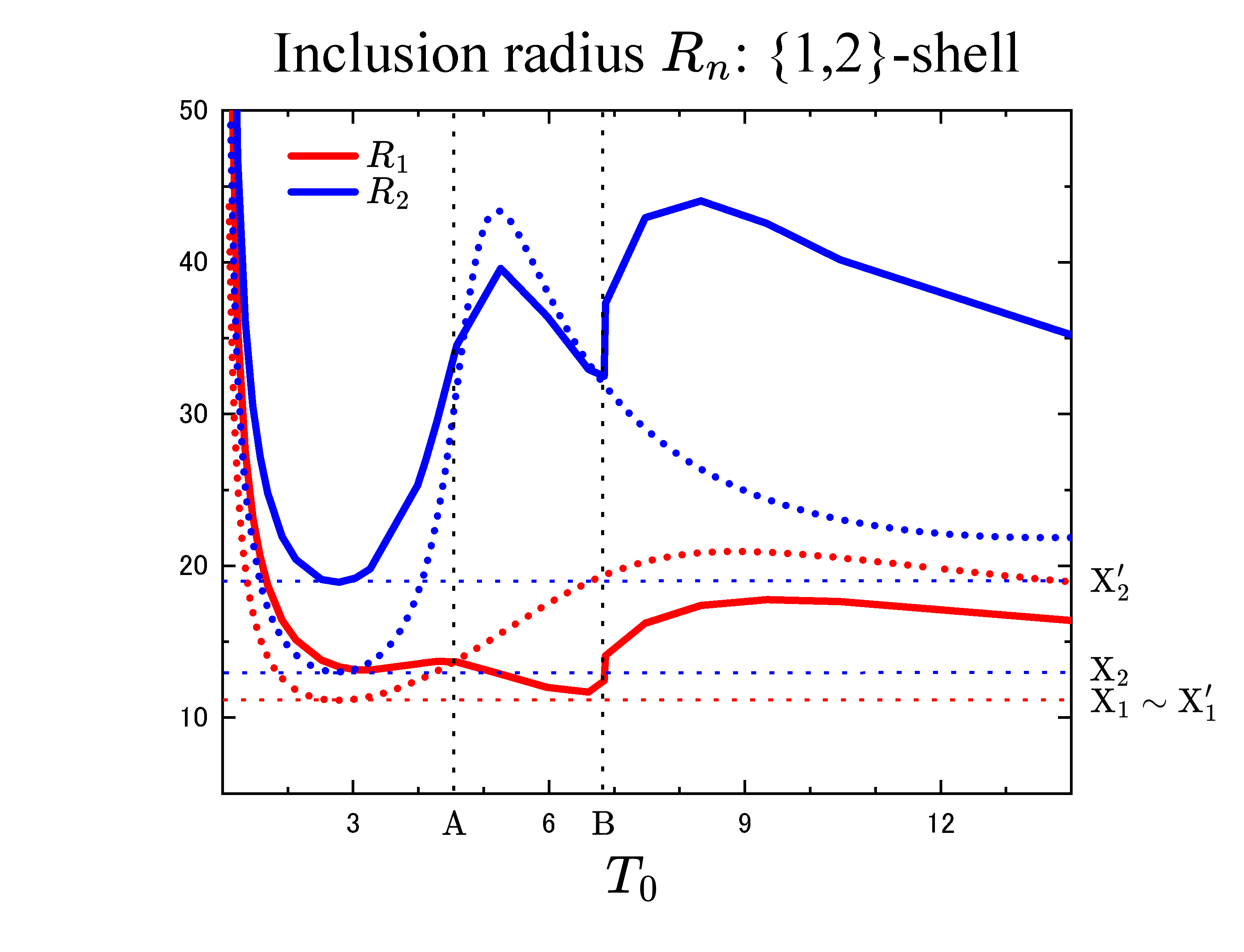}\hspace{-0.0cm}
	\includegraphics[width=0.48\linewidth]{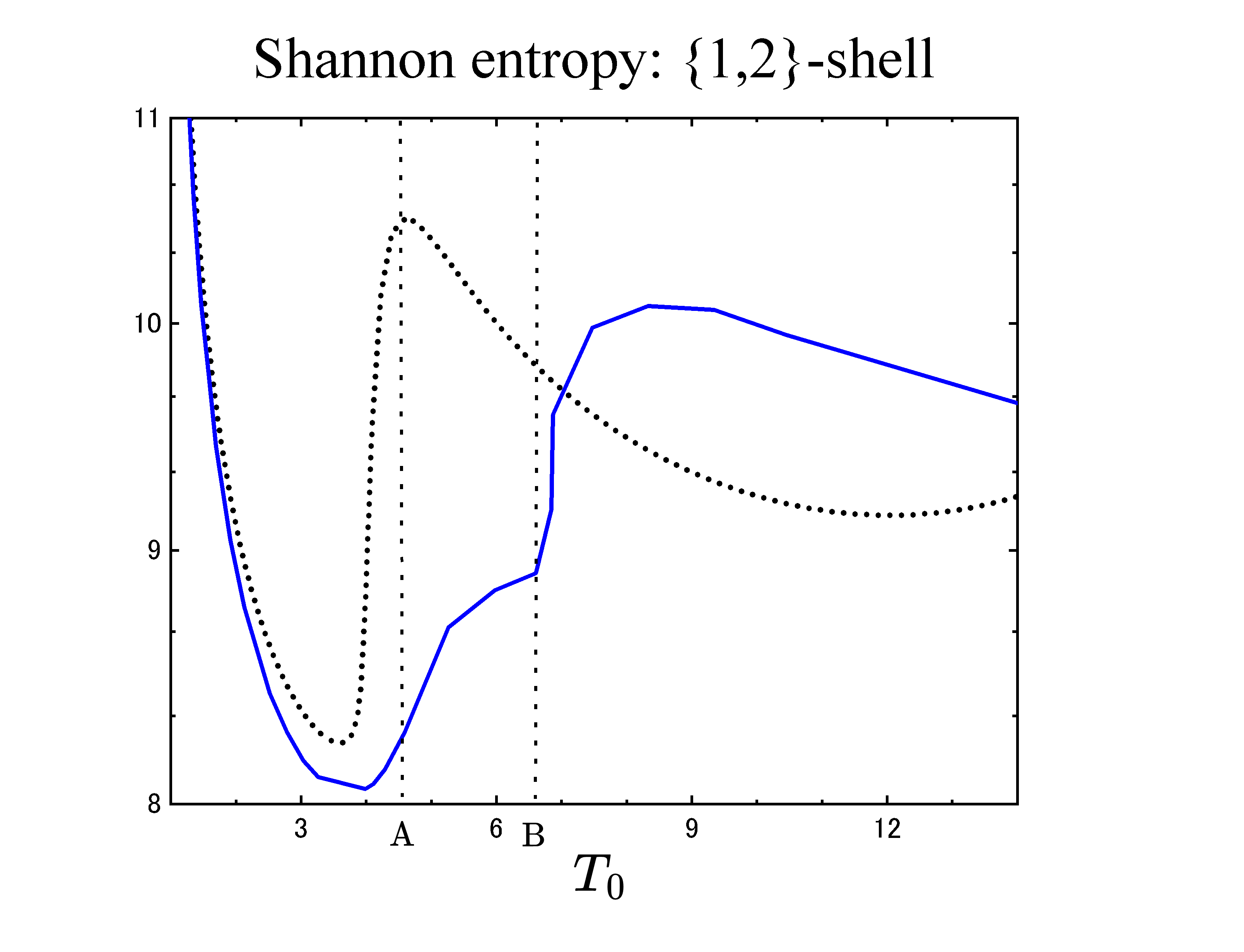}		
	\caption{\label{RS2} Inclusion radius of the $\{1,2\}$two-shell solution
	(left) and Shannon's information entropy (right). 
	As fermion number of the $n$th shell contains $2n$ fermions, $N_\textrm{f}=\{2,4\}$. 
	The dotted lines in both plots are the corresponding single-shell solutions 
	(with $N_\textrm{f}=6$ for the Shannon’s information energy). 
	In the left plot, the red and the blue lines are the solutions for $N_\textrm{f}=2$ end $4$, respectively, 
	and the points marked X$_1$,X$_2$ and X$_1'$,X$_2'$ along 
    the vertical axis are the minimum radii $R_1,R_2$ in the single-shell and two-shell solutions, respectively.  
	Points A and B along the $T_0$ indicate the redshifts at which 
	become structurally deformed. }
\end{figure*}

\begin{figure*}[t]
	\includegraphics[width=0.48\linewidth]{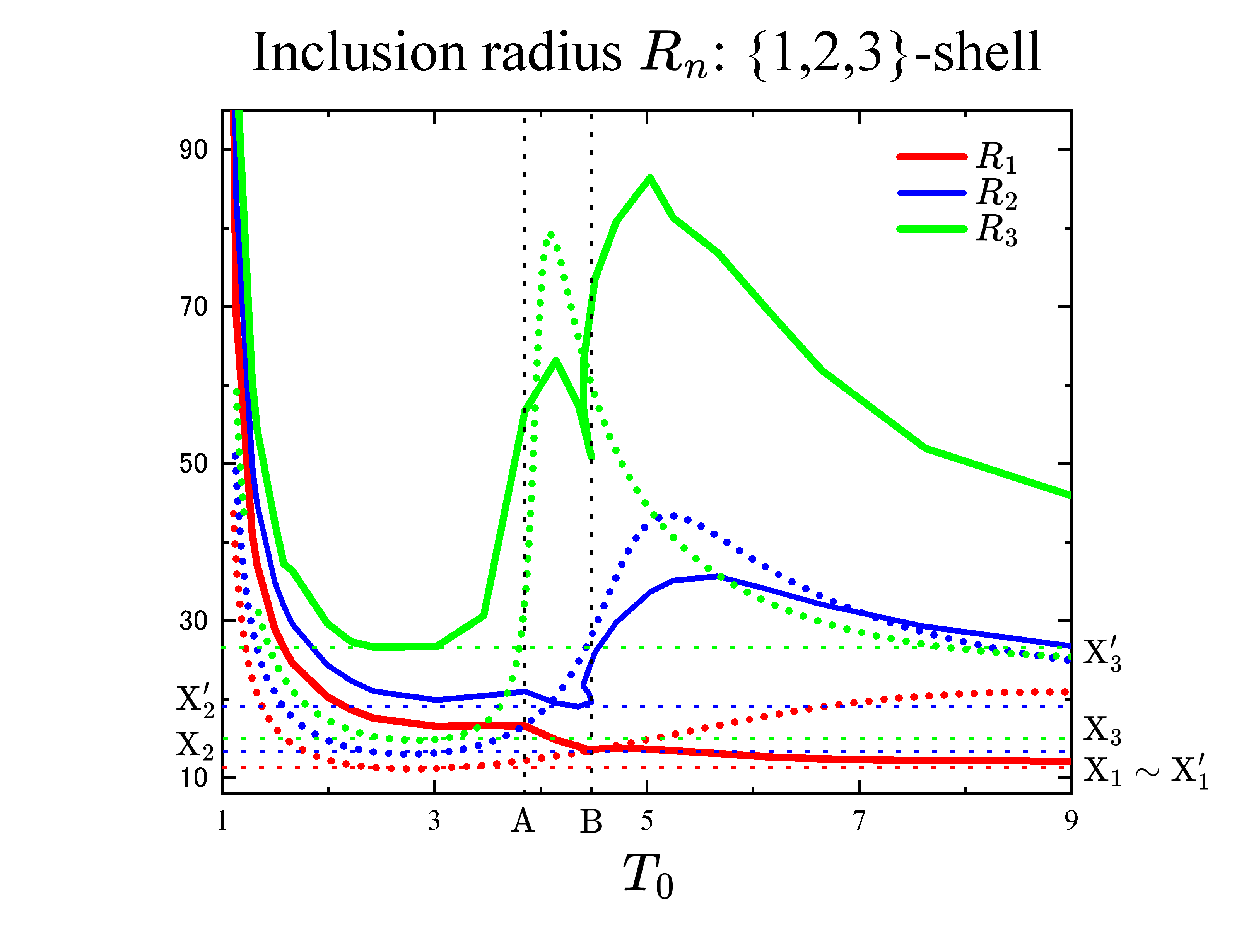}\hspace{-0.0cm}
	\includegraphics[width=0.48\linewidth]{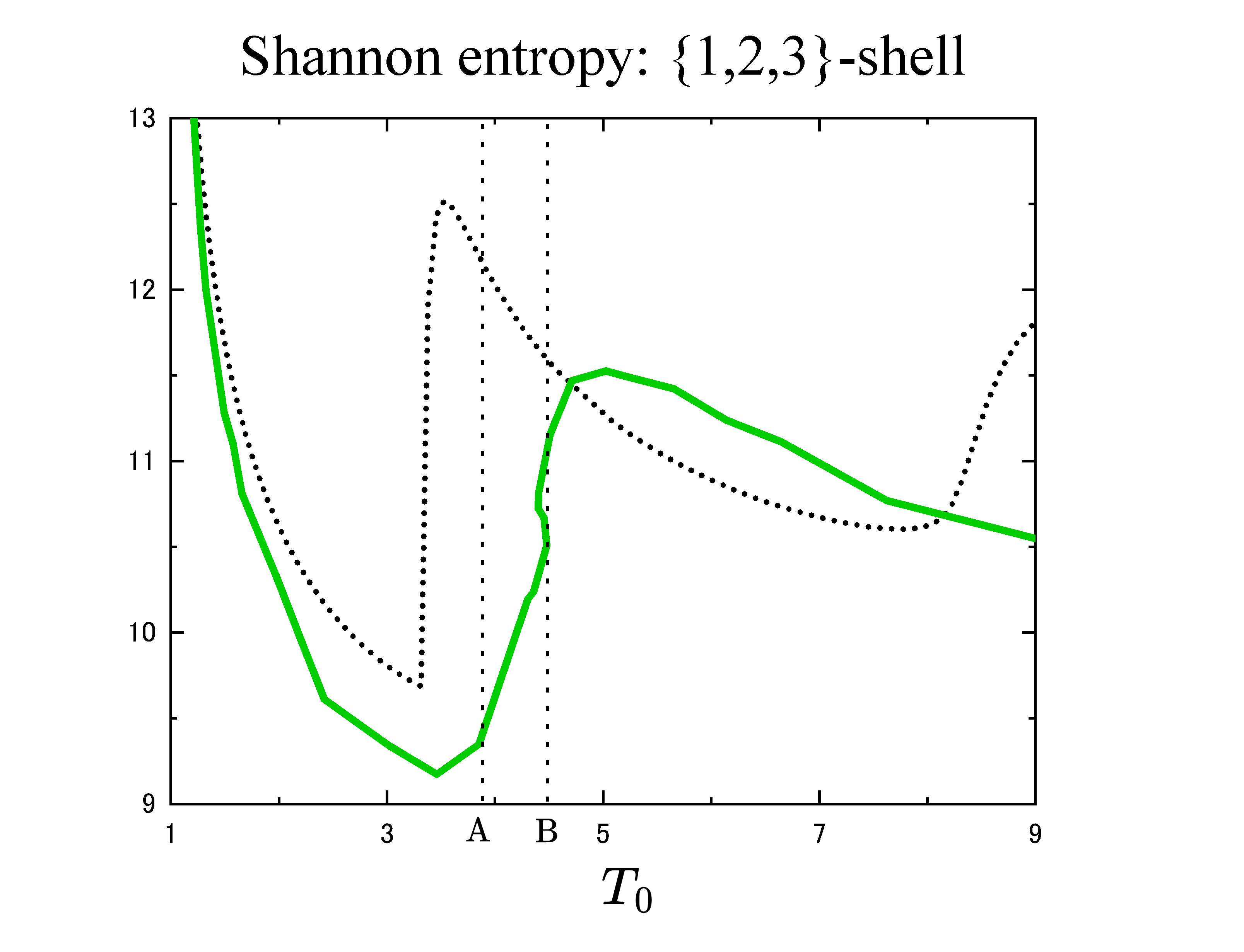}
	\caption{\label{RS3}As for Fig.\ref{RS2}, but showing the behaviors of the $\{1,2,3\}$three-shell solutions.
	In this case, $N_\textrm{f}=\{2,4,6\}$. 
	The dotted lines in both plots are the corresponding single-shell solutions (with $N_\textrm{f}=12$ for 
	the Shannon’s information energy). 
	In the left plot, the red, blue, and green lines are the solutions of $N_\textrm{f}=2,4,6$, respectively,  
	and the points marked X$_1$,X$_2$,X$_2$ and X$_1'$,X$_2'$,X$_3'$ on the vertical axis
	are the minimum radii $R_1,R_2,R_3$ in the single-shell 
	and the three-shell solutions, respectively. 
	}
\end{figure*}

\section{\label{sec:5}~Further analysis: shell fragmentation}

As shown in Figs.\ref{energy2} and \ref{energy3}, the energies decrease in the low-redshift region
and  
begin increasing from their troughs at higher redshifts. This pattern repeats after the second
minimum, reflecting the structural change in the solution; specifically,  
peak fragmentation of the shell. In fact, one can check that the energy difference
between the peak and trough is always smaller in the multishell solution than the
single-shell approximation, implying a lowering energy in the multishell solutions than 
in the single-shell solution. 
The following subsection discusses these structural changes 
in terms of pressure and entropy.

\subsection{Negative-pressure and shell deformation}

The Einstein-Dirac system satisfies semiclassical gravity theory, in which the energy-momentum 
tensor formed from 
the quantum spinor fields is subjected to normal ordering renormalization~\cite{Kain:2023jgu}.
The prominent behavior of the Einstein-Dirac system is the multishell structure 
and its fragmentation in the high-redshift region.   
The authors of Ref.\cite{Andreasson:2025uir} discussed the behavior of fragmentation 
in their Einstein-Vlasov system, a fully classical field model of gravitating fermions.  
They found a striking similarity between the multifragmentation of energy density
especially at large $N_\textrm{f}$. 
They also examined the radial pressures in both Einstein-Vlasov model 
and Einstein-Dirac models. 
Whereas only positive pressure was observed in the former model, negative pressure in 
the latter model appeared in some situations.
The authors concluded that negative pressure is an inherent quantum signature of the Einstein-Dirac system. 

The pressure in the multishell model, particularly the tail pressure in each shell, indicates 
that the shells interact to some extent while delocalizing. 
A negative radial-pressure-component indicates an attractive force between the shells 
that compacts the solution.  
The radial pressure is simply the $r$ component of the energy-momentum tensor $T^r_r$. 
In the $n$th shell, we define the radial pressure as follows: 
\begin{align}
&P_{rn}(r):=\frac{1}{2\pi r^2}\Bigl[\omega_nn T^2(\alpha_n^2+\beta_n^2)
\nonumber \\
&\hspace{2cm}-mnT (\alpha_n^2-\beta_n^2)-\frac{n^2}{2\pi r}\alpha_n\beta_n\Bigr]\,.
\end{align}
The tangential pressure in a spherically symmetric system is the $\theta$ or $\phi$ 
components $T^\theta_\theta,T^\phi_\phi$. 
In the $n$-th shell, we define
\begin{align}
P_{\perp n}(r)
=\frac{n^2\alpha_n\beta_nT}{4\pi^2r^3}\,.
\end{align}
Following~\cite{Andreasson:2025uir}, 
we plot the radial pressure on a logarithmic scale, i.e.,$\hat{P}_{rn}(r):=\log [4\pi r^2P_{rn}(r)+1]$
for visual clarity. 
Figure~\ref{1shell_NP} plots the number densities and radial pressures of the single-shell solutions  
at $T_0=3.2$ and $5.5$. 
A negative-pressure-region emerges when the shell begins fractionizing 
(and the energy increases), satisfying a structural change of the shells. 
Figure~\ref{2shell_NP} presents a similar analysis of the $\{1,2\}$-shell solutions $T_0:(2.5,7.5)$, the region of 
the shell-fragmentation onset. The spinors, number densities, and radial pressure of each spinor component are plotted at each redshift. 
During the first stage [$T_0:(2.5,6.0)$], deformation is driven by the $n=2$ shell,  which 
exhibits negative fermions pressure. 
During second stage [$T_0:(6.8,7.5)$], deformation is driven by the $n=1$ shell and 
the $n=2$ begins expanding to outer regions.  
In Fig.\ref{3shell_NP} displaying the $\{1,2,3\}$-shell solutions $T_0:(2.4,4.5)$, 
one observes negative-pressure regions in both the $n=2, 3$-shells. 

Figure~\ref{shells_high} presents the spinor, number densities and radial pressures of the 
$\{1,2\}$- and $\{1,2,3\}$-shells at a high redshift ($T_0=178.0$).
Notably, the three-shell solution (but not the two-shell solution) 
exhibits a negative outer-shell pressure. 
As seen in Figs. \ref{2shell_NP} and \ref{3shell_NP}, the outer region’s negative pressure occurs in response to the
shell fragmentation. In general, the inner shells move inside due to shell fragmentation,
causing internal compression. At the high redshift, only the three-shell model produces
shell fragmentation in the outer region, thus realizing higher compression and also lower
eigenvalues.

\subsection{Bouncing and oscillation behavior of the shell}

As evidenced above, shell fragmentation is associated with various other phenomena. 
To explicitly observe these phenomena, we newly define the shell radius $R_n$ as follows:
\begin{align}
R_n~~~\underset{\Longleftrightarrow}{{\rm def}}~~~\int_0^{R_n}(\alpha_n^2+\beta_n^2)\frac{T}{\sqrt{A}}dr\equiv 0.999
\end{align}
denoting that 99.9$\%$ of the fermions lie within $R_n$. 
Unlike the root mean square radius~\eqref{rmsr}, $R_n$ is highly sensitive to the structural change 
in the solutions. Therefore, it is a useful parameter for studying the shell-fragmentation behavior. 

To further elucidate the irregular structural changes for localizing classical field configurations, 
we apply Shannon's information entropy,  
logarithmic measure of information content in a set of discrete probabilities 
constrained to a distribution. 
Conceptualized by Shannon in 1948~\cite{Shannon1948}, 
this entropy quantifies the disorder, complexity or chaoticity of a system. 
Many recent studies in dynamical theory have adopted the Shannon entropy as a measure of
complexity or long-term stability of a system~\cite{Gleiser:2011di,Gleiser:2013mga}. 
The Shannon entropy describes the intrinsic structural change and can detects stable/unstable 
bifurcation points~\cite{Correa:2014boa,Correa:2015lla,Bazeia:2021stz,Koike:2022gfq,Feitosa:2024jtr}.
It has also been applied to gravitational theory and gravitating objects
~\cite{Gleiser:2011di,Gleiser:2015rwa,Barreto:2022ohl,Casadio:2022pla,Mvondo:1948jqw}.
The original Shannon's information entropy is defined as
\begin{align}
H=-\sum_{k=1}^np_k\log p_k\,,&n\geq1\,.
\end{align}
In spherically symmetric space-time, 
Shannon's entropy becomes
\begin{align}
&H_\textrm{D}:=\int_0^\infty p(r)\log p(r) \frac{r^2}{\sqrt{A}}dr,~~
\nonumber \\
&p(r)=\frac{1}{N_\textrm{f}r^2}\sum_{n=1}^N nT(\alpha_n^2+\beta_n^2)\,.\label{S}
\end{align}

Figure~\ref{RS2} plots the radius $R_n$ and Shannon's entropy $S_\textrm{D}$
as functions of $T_0$ for the $\{1,2\}$-shell solution. 
Also plotted are inclusion radii of the $\{1\}$- and $\{2\}$-shell models and the
Shannon entropy of the $\{3\}$-shell model. 
Structural deformation is observed at two values of $T_0$, 
(labeled A and B in Fig.\ref{RS2}) and the inclusion radii are minimized at X$_1$ and X$_2$ 
in the single-shell solutions with $N_\textrm{f} = 2$ and $N_\textrm{f} = 4$, respectively, and at 
X$_1'$ and X$_2'$ in two shell solutions with $N_\textrm{f} = 2$ and $N_\textrm{f} = 4$, respectively. 
In the low redshift region $T_0<$~A, the radii behave similarly in the single shell. 
Beyond the point $T_0=$~A, the inner shell begins to shrink and this behavior is shortly followed by  
the outer shell. 
The solution expands as it passes the point $T_0=B$, reaches its maximum, and then begins decreasing again. 
The Shannon entropy presents a somewhat different picture. 
In the region $T_0:$[A, B], the solution shrinks while the entropy monotonically grows. 
We can infer the shell structure from the minimum inclusion radii. 
In the inner radius $R_1$, the minimum X$_1$ almost coincides with that of the corresponding 
single-shell X$_1'$, suggesting similar structures of the two-shell models, at least within the inner shell.  
The outer shell is more rigid to compression because the minimum X$_2'$ of the outer radius 
$R_2$ in the two-shell model exceeds that of the single shell. 
Obviously, compression is prevented by a degeneracy pressure exerted by the fermions in the interior. 
However, at redshifts $T_0>$~A, this picture is apparently broken because the order of the inner radius swaps and 
the outer shell suddenly and vastly expands at $T_0>$~B.
This behavior requires a different interpretation. 

The entropy describes the complexity of a localizing object and an entropy change
indicates that a structural deformation in the solution. 
The sudden explosion of the solution at $T_0=$~B indicates a drastic structural change. 
More specifically, the entropy's growth in the two-shell solution comprises two consecutive steps, deviating from 
the entropy behavior of the single shell solution. The second step involves a larger phase transition 
and a change from the structure at low $T_0$. 
The order-swapping of the energy in Fig.\ref{energy2} originates from the rise in the eigenvalues $\omega_n$ with 
increasing radius $R_n$.

The three-shell solution presents more complicated behavior. Figure~\ref{RS3} shows the inclusion radius and the 
Shannon entropy of the $\{1,2,3\}$-shell solution. In the $T_0:$~[A, B], large structural changes occur and at 
least two subsequent changes appear in the Shannon entropy. Within the inner radius $R_1$, 
the minimum X$_1$ almost coincides with that of the corresponding single shell X$_1'$. 
Within the outer shells, 
the minima X$_2'$, X$_3'$ are higher than those within the single shell solution. 
The difference between the two-shell model is more evident in the outer shell.

\section{\label{sec:6}~Summary and outlook}

We have developed a multishell formulation of the Einstein-Dirac system and 
obtained the ground states of closed shells with the fermion numbers $N_\textrm{f}=6,12,20$.
Shell fragmentation at high redshift was the most prominent feature of the many-fermions
in the single-shell approximation of the Einstein-Dirac system. 
In the multishell model, the outer shell moves outward while the inner shell moves 
inward, creating a high-density phase. 
The $\{1,2,3\}$-shell model yields almost lower energy and eigenvalues than the 
corresponding single-shell approximation. From this result, we conclude that the solution 
is a ground state of the model with
fermion number $N_\textrm{f}=12$. 
In contrast, the energy in the $\{1,2\}$-shell model is comparable to or 
slightly higher than the energy in the single-shell model at the high redshifts ($T_0>10$), suggesting
that properties of the energy spectrum differ from those of the conventional shell-like model.  
The radial pressure provides information on the structural deformations of an object. We found that 
negative pressure emerges when the shell is fragmented. We also introduced two important
parameters---inclusion radius and the Shannon's informational entropy---for understanding the 
behavior of shell fragmentation. Two successive phases of 
structural changes were observed at shell-bouncing. 
Unlike the two-shell solution, the three-shell solution exhibited a negative outer-shell pressure 
in high redshift region. The three-shell solution yielded overlap between the shells, higher compression and
lower eigenvalues at high redshift. 

The following open problems should be resolved in future research. 
The present system admits many excited states, especially in the nodal solutions of the spinors. 
These states
correspond to negative-parity states and provide rich physical information. 
Many authors have models coupled with electromagnetism or Yang-Mills fields. 
The competition between the repulsive electromagnetic force and the gravitational force warrants
a new phase diagram for describing stable/unstable branches in terms of sign of some of these 
interactions. 
In this paper, we studied shell models with four or fewer shells. Therefore, fermion number was
$N_\textrm{f}=20$. 
Similar technological problems persist in the many-nucleon traditional shell model, 
but are recently being overcome with  
deep learning algorithms~\cite{Negoita:2018kgi,Mazur:2024psp,Sharypov:2025hiz}. 
Adopting such methods, we hope to solve our problem with larger numbers of shell
in our subsequent papers. 
\vspace{1cm}

\noindent {\bf Acknowledgments} 
The authors thank Yakov Shnir, Atsushi Nakamula, Ryu Sasaki, Filip Blaschke,
and Pawe\l~Klimas for their practical advice and valuable comments. 
N.S. is supported in part by JSPS KAKENHI Grant No. JP23K02794.

\vspace{0.5cm}
\appendix

\section{Shannon entropy}

Appendix derives the Shannons' informational entropy of a continuous function $p(r)$. 
The original Shannon entropy was defined for discrete probability distributions~\cite{Shannon1948}. Let
\begin{align}
\mathbb{D}_n=\left\{\left(p_1,p_2\cdots,p_n\right)|p_k\geq0,k=1,2,\cdots,n,\sum_{k=1}^np_k=1\right\},\nonumber\\
n\geq1
\end{align}
denote a set of $n-$component discrete probability distributions. The Shannon entropy $H:\mathbb{D}_n\to\mathbb{R}$ is defined 
as follows:
\begin{align}
H(p_1,p_2,\cdots,p_n)&=-\sum_{k=1}^np_k\log p_k,&n\geq1
\end{align}
with $0\log0=0$ by convention.  
As in Ref.\cite{Gleiser:2018jpd}, we can directly generalize the entropy to a continuous function
by setting $p_k= p(k)dk$ where $p(k)$ is a constant in the period $k\sim k+dk$ 
\begin{align}
H_D&=-\lim_{dk\to0}\sum_{k\in \mathbb{D}}p(k)dk\log p(k)dk\nonumber\\
&=-\lim_{dk\to0}\sum_{k\in \mathbb{D}} p(k)dk\log p(k)-\lim_{dk\to0}\sum_{k\in \mathbb{D}} p(k)dk\log dk\nonumber\\
&=-\int dkp(k)-\lim_{dk\to0}\log dk.\label{original_S}
\end{align}
In the last line, we have used $\sum_k p(k)dk=1$.
The last term of \eqref{original_S} is a logarithmically diverging term caused by moving the continuum limit. 
The continuum limit can be removed by introducing an appropriate regularization cutoff.  
We adopt the first term only in \eqref{original_S} as the Shannon entropy of a the continuous function. 
In the present analysis, we define the probability density on a coarse scale as follows:
\begin{align}
&p(k)\to \dfrac{\displaystyle\sum_n\displaystyle\sum_{k_n=-j_n}^{j_n}|\Psi_{j_nk_n}(x)|^2}
{\displaystyle\sum_{n}\int\displaystyle\sum_{k_n=-j_n}^{j_n}|\Psi_{j_nk_n (x)}|^2\sqrt{-\textrm{det}(g_{ij})}d^3x}\,,
\\
&dk\to \sqrt{-\textrm{det}(g_{ij})}d^3x\,.
\end{align} 

\vspace{0.5cm}
\section*{References}

\bibliography{Dirac}

\end{document}